\documentclass{jfm}
\usepackage{graphicx}
\usepackage{epstopdf, epsfig}
\usepackage{amsmath}
\usepackage{mathtools}
\usepackage{todonotes}
\usepackage{textgreek}

\usepackage{soul}
\usepackage{cancel}
\usepackage{mathrsfs}
\usepackage{newtxtext}
\usepackage{amssymb}
\usepackage{bm}
\usepackage{booktabs, multirow} % for borders and merged ranges
\usepackage{float}
\usepackage{placeins}
\usepackage{url}

\def\ee{{\rm e}}
\def\ii{{\rm i}}

\newcommand\blue{\textcolor{blue}}

\usepackage{fix-cm} % not needed if not using Computer Modern
\DeclareMathAlphabet{\mathsfit}{T1}{\sfdefault}{\mddefault}{\sldefault}
\SetMathAlphabet{\mathsfit}{bold}{T1}{\sfdefault}{\bfdefault}{\sldefault}

\shorttitle{Linear stability of an oceanic front}
\shortauthor{S. Kar, R. Barkan \& J. R. Taylor}

\title{Linear stability of an oceanic front at finite Rossby number}
\author{Subhajit Kar\aff{1}
\corresp{\email{subhajitkar@mail.tau.ac.il}},
 Roy Barkan\aff{1,2}
 \and John R. Taylor \aff{3}
}
\affiliation{\aff{1} Porter School of the Environment and Earth Sciences, Tel Aviv University, Tel Aviv 69978, Israel
\aff{2}Department of Atmospheric and Oceanic Sciences, University of California, Los Angeles, CA, USA
\aff{3}DAMTP, University of Cambridge, Centre for Mathematical Sciences, Wilberforce Road, Cambridge CB3 0WA, UK.
}

\begin{document}

\maketitle

\begin{abstract}
Submesoscale currents in the ocean’s mixed layer (ML), consisting of fronts, eddies, and filaments, are characterized by $\mathcal{O}(1)$ Rossby $(Ro)$ and Richardson $(Ri)$ numbers. These currents play a crucial role in mediating vertical exchange between the surface and ocean interior and in facilitating cross-scale energy transfers. Despite a growing understanding of their generation mechanisms and energy pathways, two fundamental questions remain unresolved \--  how does a finite $Ro$ modify the dynamics of ML instabilities, and what mechanisms are responsible for ML frontal arrest when $Ro \sim \mathcal{O}(1)$. In this study, we address these questions through a linear stability analysis of a two-dimensional, geostrophically adjusted oceanic front based on the analytical model of \cite{ou1984geostrophic}, which allows systematic exploration across a range of $Ro$. In the low $Ro$, $Ri \sim \mathcal{O}(1)$ regime, the most unstable mode is that of baroclinic instability, with the buoyancy flux serving as the primary source of perturbation kinetic energy. As $Ro$ increases, the dominant instability becomes an inertia-critical layer type, characterized by a resonant interaction between a Rossby wave and an inertia-gravity wave. In the $Ro \sim \mathcal{O}(1)$ regime, the shear production terms become comparable to the buoyancy flux term and even dominate in the region where the adjusted front is strongest. Our results suggest that shear production should be included in parameterizations of ML instabilities.     
\end{abstract}

% \begin{keywords}
% Authors should not enter keywords on the manuscript, as these must be chosen by the author during the online submission process and will then be added during the typesetting process (see http://journals.cambridge.org/data/\linebreak[3]relatedlink/jfm-\linebreak[3]keywords.pdf for the full list)
% \end{keywords}

\section{Introduction}
\label{intro}
Surface submesoscale currents, comprising mixed layer eddies, fronts, and filaments, are pervasive features in the mixed layer (ML) of the upper ocean, exhibiting horizontal scales of $\mathcal{O}(0.1$–$10)$ km and temporal scales ranging from hours to days. These submesoscale currents are characterized by $\mathcal{O}(1)$ Rossby number $Ro\,(=U_0/fL_0)$ and Richardson number $Ri\,(=N^2 H^2/U_0^2)$, where $U_0$ is a horizontal velocity scale, $L_0$ is a horizontal length scale, $H$ is a vertical length scale, $f$ is the Coriolis parameter, and $N$ is the buoyancy frequency \citep{mcwilliams2016submesoscale}. This distinguishes them from oceanic mesoscale eddies, which have horizontal scales of $\mathcal{O}(10$–$100)$ km and a timescale of many days, characterized by $Ro\ll 1$ and $ Ri\gg 1$. Owing to their large vertical velocities, submesoscale currents are instrumental in mediating the exchange of momentum, heat, and tracers between the surface and the ocean interior, thereby modulating a broad spectrum of physical and biogeochemical processes \citep{thomas2008submesoscale, mahadevan2016impact, taylor2023submesoscale}.

Submesoscale currents are generated through several processes, including baroclinic mixed layer instabilities \citep[MLIs;][]{boccaletti2007mixed,fox2008parameterization}, strain driven frontogenesis \citep{hoskins1972atmospheric}, and frontogenesis induced by boundary layer turbulence \citep{gula2014submesoscale,mcwilliams2015filament, dauhajre2025vertical}. These mechanisms contribute to the restratification of the ML by releasing the available potential energy (APE) stored in horizontal density gradients, and to bidirectional cross-scale kinetic energy (KE) fluxes \citep{capet2008mesoscale3, balwada2022direct, srinivasan2023forward}. Specifically, finite-amplitude MLIs develop into mixed layer eddies (MLEs), which have been shown to exhibit an inverse KE cascade \citep{fox2008parameterization}. Consequently, KE is transferred upscale and can energize mesoscale eddies \citep{klein2019ocean, schubert2020submesoscale, srinivasan2023forward}. In turn, mesoscale strain and boundary layer turbulence can initiate submesoscale frontogenesis \citep{barkan2019role}, thereby driving a forward KE flux that can deplete mesoscale KE \citep{srinivasan2023forward, yu2024intensification}. The pathways of these forward KE fluxes to dissipation are ultimately determined by the processes that lead to frontal arrest. Under suitable forcing conditions, symmetric instability can emerge as the arrest mechanism \citep{thomas2005destruction}; otherwise, horizontal shear instability becomes a likely candidate  \citep{sullivan2018frontogenesis}.

% Historically, the distinction between MLIs and the classical baroclinic instability \citep[BCI,][]{eady1949long} has largely been associated with the Richardson number, which is assumed $\mathcal{O}(1)$ (figure \ref{fig:MLI_Ro_Ri}). 

\begin{figure}
    \centering
    \includegraphics[width=0.6\linewidth]{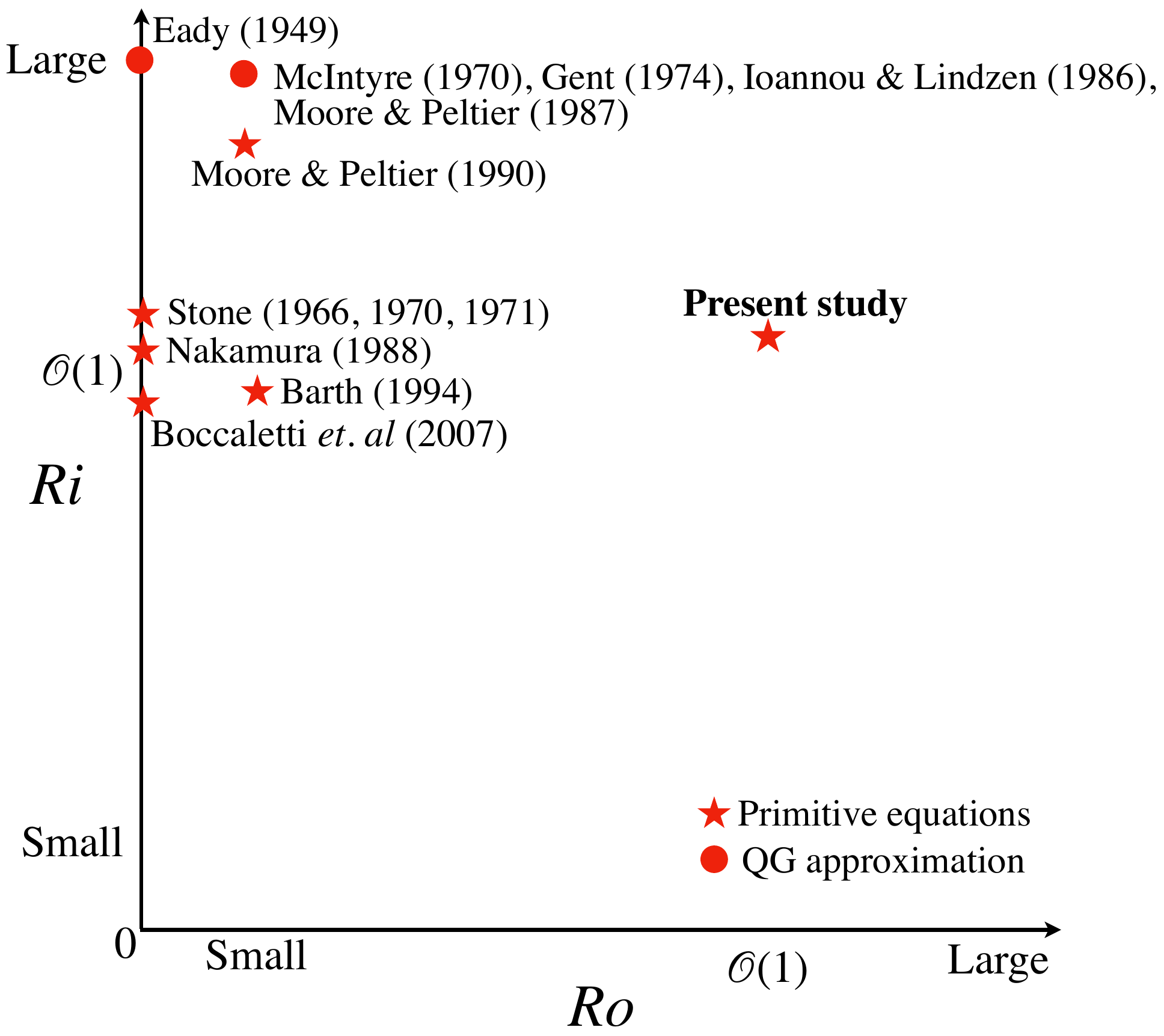}
    \caption{The parameter space examined by previous linear stability studies of {baroclinic front configurations}, shown schematically as a function of Rossby ($Ro$) and Richardson $(Ri)$ numbers {of the basic state.} Here the Rossby number is defined in terms of the vertical relative vorticity of the basic state. {In this study, the basic state allows us to specifically investigate the regime where $Ro \sim \mathcal{O}(1)$ and $Ri \sim \mathcal{O}(1)$.}%, where $Ro$ and $Ri$ in the present study is defined as the bulk Rossby and Richardson numbers ((\ref{def_Ro}) and (\ref{def_Ri})). } }
    }
    \label{fig:MLI_Ro_Ri}
\end{figure}

Historically, the distinction between MLIs and classical baroclinic instability \citep[BCI,][]{eady1949long} has been framed primarily in terms of the Richardson number, $Ri$, with MLIs typically associated with $Ri \sim \mathcal{O}(1)$ (figure \ref{fig:MLI_Ro_Ri}).
Most notably, \cite{stone1966non,stone1970non,stone1971baroclinic} investigated non-geostrophic effects on BCI using Eady’s framework. He found that as the $Ri$ decreases, the wavelength of the most unstable mode increases while the growth rate diminishes relative to predictions from the quasigeostrophic (QG) approximation. The effects of horizontal shear on BCI have been studied in the QG regime by \citet{mcintyre1970non} and \citet{gent1974baroclinic}. They demonstrated that horizontal shear substantially alters the classical Eady problem by producing a counter-gradient horizontal momentum flux that stabilizes the BCI \-— a mechanism referred to as the barotropic governor \citep{james1987suppression}. In addition, cross-front variations in the mean flow were shown to spatially confine BCI modes in the cross-front direction \citep{ioannou1986baroclinic, moore1987cyclogenesis}. 

\citet{moore1990nonseparable} and \citet{barth1994short} investigated the stability of a 2D front using the primitive equations in the $Ro\ll 1, Ri\gg 1$ regime. They identified two distinct instability modes \-- a long-wave mode and a short-wave mode. The long-wave instability mode closely resembles the classical Eady-type BCI and arises from the phase-locking of two counter-propagating Rossby waves \citep{hoskins1985use}. In contrast, the short-wave instability, absent in the QG approximation, can be viewed as a continuation of BCI, wherein one Rossby wave is replaced by an inertia-gravity wave (IGW). This instability mechanism involves a resonant interaction between a Rossby wave and an IGW and is commonly referred to as inertia-critical layer (ICL) instability \citep{stone1970non,nakamura1988scale}.

The typical length scale of MLIs is estimated to lie within the range $4 \le L_\text{MLI}/R \le 6$ \citep{eldevik2002spiral,ozgokmen2011large}, where $L_\text{MLI} \approx 4R\sqrt{1 + Ri^{-1}}$ represents the wavelength of the most unstable MLI mode \citep{stone1966non}, $R$ is the ML deformation radius, and $H$ denotes the ML depth. Using the upper bound of this range (corresponding to $Ri = 0.8$), \cite{dong2020scale} provided a global estimate of the MLI length scale in the ocean, finding a typical value of approximately $6$ km at mid-latitudes. High-resolution realistic ocean simulations capable of resolving these scales indicate that the associated local Rossby number in such solutions is typically $\mathcal{O}(1)$ \citep[e.g.,][]{capet2008mesoscale1, barkan2017submesoscale}, implying that horizontal shear may have important effects on MLI dynamics. 
Furthermore, \citet{bodner2023modifying} recently revised the MLI-induced restratification parametrization originally developed by \citet{fox2008parameterization}, incorporating an arrested frontal length scale derived from the theory of turbulent thermal wind \citep{mcwilliams2015filament}. This suggests that MLIs are expected to also occur at spatial scales smaller than the ML deformation radius $R$, where the local Rossby number may be even larger and is expected to influence  the growth rate and energetics of the instability. In addition, at these smaller scales, the finite Rossby number may potentially have important implications for frontal arrest mechanisms.

The preceding discussion motivates two fundamental open questions:
\vspace{0.1 cm}
\begin{enumerate}
\item how does a finite $Ro$ affect the characteristics of  MLI? 
\item what instability mechanisms can arrest submesoscale frontogenesis when $Ro \sim \mathcal{O}(1)$? 
\end{enumerate}
\vspace{0.1 cm}
In this paper, we address these questions through a bi-global linear stability analysis \citep{theofilis2011global} of a geostrophically adjusted front. 
The classical frontal geostrophic adjustment problem describes the evolution of an isolated front, starting from a state of rest \citep{rossby1937mutual}. At early times, a vertically sheared cross-front circulation develops in response to the difference between the hydrostatic pressure on each side of the front. As the flow evolves, the cross-front circulation drives frontogenesis, before the front adjusts to a geostrophically balanced state. 

%For sufficiently strong fronts in the absence of viscosity, no adjusted state exists \citep{ou1984geostrophic} \blue{(is this really the right paper to cite here? should HB72, or something like that be better?)}, and a singularity develops in finite time \citep{blumenwu}.

We consider a basic state based on the analytical solution of a geostrophically adjusted ML front as described by \citet[][{herein after Ou84}]{ou1984geostrophic}, enabling us to systematically explore stability properties across a range of $Ro$ values that characterize the strength of the front. When $Ro$ exceeds a critical value, $Ro>Ro_c$, the solutions exhibit a discontinuity, which makes the stability problem ill-posed. By considering smaller Rossby numbers with $Ro<Ro_c$, we seek instabilities that may act to equilibrate frontogenesis in the full time-dependent geostrophic adjustment problem. 

We demonstrate that in the $Ro\ll1, Ri\sim \mathcal{O}(1)$ regime, the adjusted front is unstable to BCI, and the growth rate of the most unstable mode closely matches the analytical solution of \citet[][herein after S71]{stone1971baroclinic}, with the buoyancy flux acting as the primary source of perturbation KE. 
In the oceanic submesoscale regime, characterized by $Ro \sim \mathcal{O}(1)$ and $Ri \sim \mathcal{O}(1)$, the most unstable growth rate remains comparable to that of S71, but the corresponding mode transitions to the ICL instability mode described above. In this regime, the buoyancy flux and the shear production terms contribute equally to the growth of the perturbation KE.

%is driven primarily by buoyancy flux, in combination with horizontal- and vertical-shear production.}

The paper is organized as follows. In \S \ref{model_setup}, we discuss the basic-state configuration (\S \ref{mean_flow}) and the linear stability analysis (\S \ref{perturb}). Results of the stability analyses are shown in \S \ref{results}. In \S \ref{discussion}, we discuss the results and draw connections to the two open questions above. Finally, in \S \ref{summary}, we summarize our findings.

% {The details of the numerical setup are provided in \S \ref{numerical_model}}.
% {In \S \ref{results}, we discuss the evolution of the mean flow and compare the numerical solution with 2D semi-analytical frontogenesis solutions.}
% A detailed analysis of the front-IW energy exchanges is shown in \S \ref{sec_energetics}. Finally, in \S \ref{summary}, we summarize our findings and draw connections to realistic ocean scenarios. 

% The presence of a seasonal cycle in the submesoscale flows suggests that the BCI can play an important role in the dynamics of the upper ocean.

% There are different generation mechanisms leading to the formation of these structure.
% These are typically generated by straining motions provided by the large-scale mesoscale eddies (\citep{hoskins1972atmospheric}). 

\section{Problem configuration}
\label{model_setup}
We perform a bi-global stability analysis of a
$2$D front in the $(y,z)$ plane that is invariant in the along-front ($x$) direction.
 The dynamics are governed by the Boussinesq equations of motion for a rotating fluid under the $f$-plane approximation. To non-dimensionalize the governing equations, we follow the scaling proposed by Ou84. The length and time scales are,
\begin{subequations}
\begin{gather}
\label{len_scale}
    {x} = R \hat{x}, 
    \,\,\,\
    {y} = R \hat{y}, \,\,\,\
    {z} =  H \hat{z}, \,\,\,\
    {t} = \frac{1}{f} \hat{t}, 
    \tag{\theequation $a-c$}
\end{gather}
\end{subequations}
where the `hat' describes a non-dimensional quantity, 
$H$ is the ML depth, $R = \sqrt{\Delta B H}/f$ with $\Delta B$ is the buoyancy anomaly across the front, with the buoyancy $B=-g\rho/\rho_0$ ($\rho$ is the flow density perturbation relative to the reference density $\rho_0$, and $g$ is the gravitational acceleration). The corresponding scales of the other flow fields are
\begin{subequations}
\label{flow_scale1}
\begin{gather}
\label{flow_scale}
    {U} = fR\widehat{U},
    \,\,\,\
    {V} = f R\widehat{V},
    \,\,\,\
    {W} =  f H \widehat{W}, \,\,\,\
    {P} = f^2 R^2 \widehat{P}, \,\,\,
    {B} = (f^2 R^2/H) \widehat{B},
    \tag{\theequation $a-d$}
\end{gather}
\end{subequations}
where $\widehat{U}, \widehat{V}$, and $\widehat{W}$ denote the non-dimensional velocity components in the $x$, $y$, and $z$ directions, respectively; $\widehat{P}$ is the non-dimensional pressure, and $\widehat{B}$ is the non-dimensional buoyancy. To simplify notation, we omit the `hat' symbol in the following analysis unless explicitly stated.

\subsection{Basic state}
\label{mean_flow}
The basic state is defined following
Ou84, in which an initially motionless fluid with a lateral buoyancy gradient is geostrophically adjusted toward a balanced state, i.e., the along-front mean flow is in geostrophic balance with the lateral pressure gradient. For an imposed buoyancy profile $B(\eta)$,  
Ou84 derived the following solution for the along-front velocity $U$ of the adjusted state, 
\begin{subequations}
\label{ou1984_sol}
\begin{align}
    y &= \Big(\frac{1}{2} - z \Big)  \frac{dB}{d \eta} + \eta, 
\\
    z &= \Big( \frac{d^2 B}{d \eta^2} \Big)^{-1}
    \Bigg[1 + \frac{1}{2} \frac{d^2 B}{d \eta^2} - \sqrt{\Big(1+\frac{1}{2}\frac{d^2 B}{d \eta^2} \Big)^2 - 2 \frac{d^2 B}{d \eta^2} \xi} \Bigg],
\\ 
 U(y,z) &= \Big(\frac{1}{2} - z \Big)  \frac{d B}{d \eta},
\end{align}  
\end{subequations}
where $(\eta, \xi)$ denote the initial coordinates of a fluid particle that ends at  $(y,z)$ after geostrophic adjustment. 
In this study, we use 
\begin{align}
\label{buoy_front}
    B(\eta) = -\frac{1}{2} \tanh{(\beta \eta)},
\end{align}
where $\beta$ is a free parameter that measures the steepness of the initial buoyancy profile. Since the fluid is initially motionless and the prescribed buoyancy distribution (\ref{buoy_front}) is depth-independent, the initial potential vorticity (PV) is zero.
The geostrophic adjustment process conserves PV materially, and therefore, the PV of the adjusted front also remains zero. This is a physically relevant choice, as low PV typically characterizes oceanic ML fronts.

%%%

% To quantify the dynamics in terms of frontal Rossby number $Ro_\text{front}$ instead of $\beta$, we calculated it based on the definition in \cite{wienkers2021influence}, which is defined as 
% \begin{align}
%     Ro_\text{front} = \frac{1}{\delta_\text{front}}\underbrace{\frac{1}{f^2} \frac{f^2 R}{H}{|\partial_\eta B}|}_{\Gamma=M^2/f^2} = \frac{\beta^2}{4} ,
% \end{align}
% where $\delta_\text{front} = (L_f R)/H$ is the spacing between two fronts (figure \ref{fig:Ro_Ri}$(a,b)$). The variable $\Gamma$ denotes frontal strength, is defined as the dimensional horizontal buoyancy normalized by $f^2$.

To characterize the frontal dynamics, we introduce a bulk Rossby number, $Ro$, defined in terms of the basic state velocity, $U_0$, and frontal width, $L_0$,
\begin{align}
\label{def_Ro}
    Ro = \frac{U_0}{fL_0}= \frac{\beta^2}{4},
\end{align}
where $U_0=fR\beta/2$ (e.g., (\ref{ou1984_sol}$c$)) and the typical dimensional cross-front length scale is $L_0=2R /\beta$ as noted above. Therefore, the non-dimensional frontal width, $\beta$, is linked with the Rossby number, and the dimensional cross-frontal width is $Ro^{-1/2} R$). The bulk Richardson number of the front is defined as
\begin{align}
\label{def_Ri}
    Ri = \frac{N^2H^2}{U_0^2} = 1,
\end{align}
where $N^2 = f^2 R^2/H^2 |{dy}/{dz}| {\beta}/{2}$ and 
$|{dy}/{dz}|=\beta/2$ is the slope of the adjusted isopycnals  (using \ref{ou1984_sol}$a$).
{This $Ri$ number definition agrees with the local value computed from the frontal solutions ($Ri_\ell \equiv \partial_z B / (\partial_z U)^2 \sim \mathcal{O}(1)$; figures \ref{fig:Ro_Ri}$(c,d)$). }
%The value of $Ri \sim \mathcal{O}(1)$ is consistent with the frontal solution (figures \ref{fig:Ro_Ri}$(b,d)$).

%The solution described by (\ref{ou1984_sol}$a-c$) breaks down when isopycnals intersect. { This occurs when $\partial_y \eta =0$. According to (\ref{ou1984_sol}$(a)$), this happens when $|\partial_\eta ^2 B|>2$ (\red{is it 2 or 2.28? I assume the singularity occurs because of the intersecting isopycnals? }) at the top and bottom surfaces of the domain. Non-singular solutions to (\ref{ou1984_sol}$a-c$) require that $\beta \lesssim 2.28$.}
%We therefore perform a linear stability analysis of the above basic state with $0.1\leq\beta\leq 2$.
%In this study, we performed a stability analysis of the mean flow with a maximum value of $\beta$ set to $2$.

\begin{figure}
    \centering
    \includegraphics[width=\textwidth]{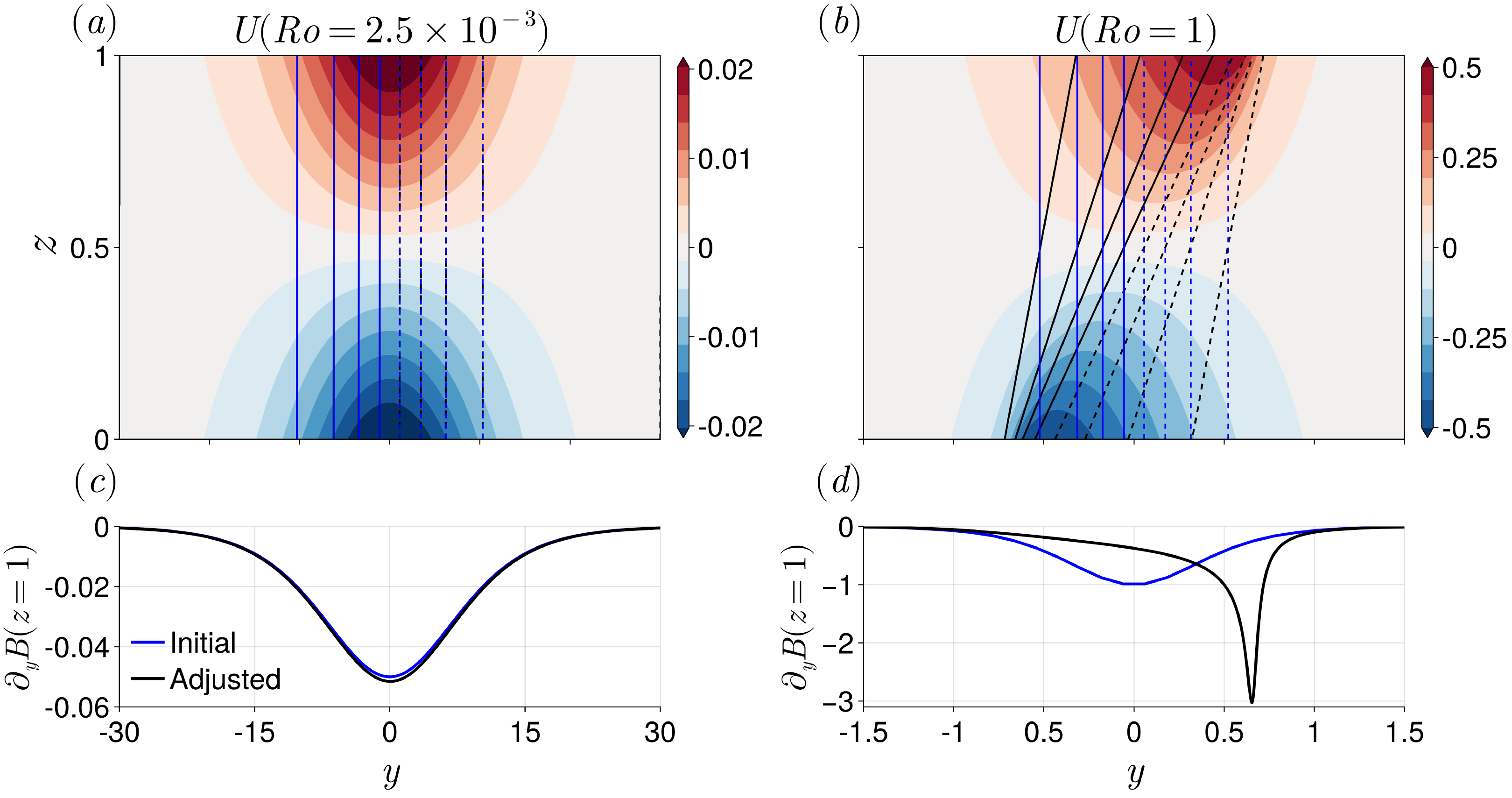}
    \caption{The solution of geostrophically adjusted ML front  described by (\ref{ou1984_sol}$a-c$) for $(a)$ $Ro=2.5\times10^{-3}$ and $(b)$ $Ro=1$. The blue (black) lines show the initial (adjusted) buoyancy contours with a contour interval of $B=0.11$. Solid (dashed) lines denote positive (negative) values. Panels (c) and (d) show the initial (blue) and adjusted (black) horizontal buoyancy gradient $\partial_y B$ at the top surface $(z=1)$ corresponding to $Ro=2.5\times10^{-3}$ and $Ro=1$, respectively.     }
    \label{fig:mean_flow}
\end{figure}

\begin{figure}
    \centering
    \includegraphics[width=\textwidth]{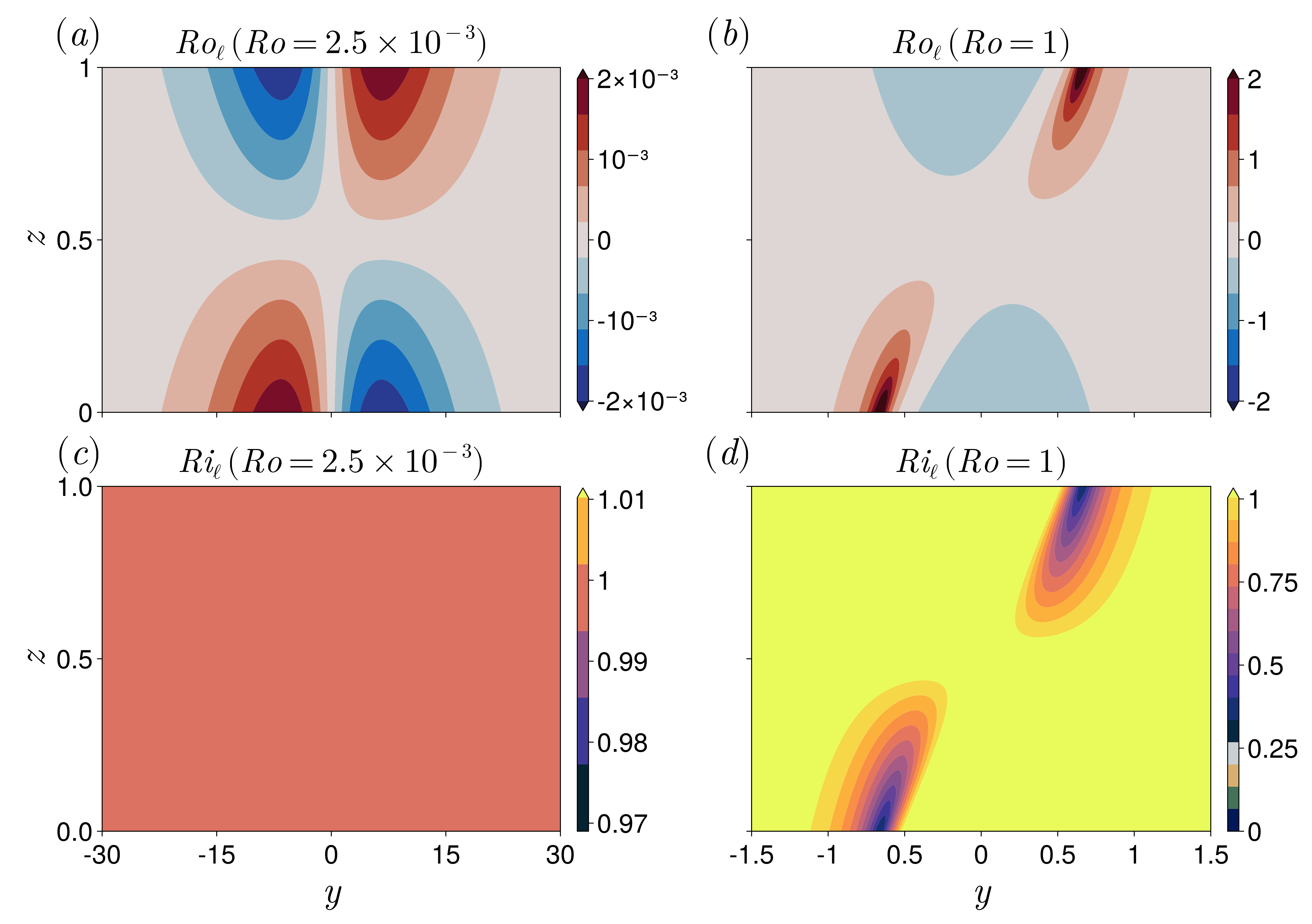}
    \caption{$(a,b)$ The local Rossby number $Ro_\ell=-\partial_y U$ and $(c,d)$ the local Richardson number $Ri_\ell=\partial_z B/(\partial_z U)^2$ of the mean flow shown in figures \ref{fig:mean_flow}$(a,b)$. 
    In both cases, $Ri_l > 1/4$, therefore, the vertical shear of the mean flow is stable with respect to Miles-Howard criterion \citep{miles1961stability}.}
    \label{fig:Ro_Ri}
\end{figure}

For small $Ro$ values, frontogenesis is relatively weak, resulting in minimal differences between the initial and adjusted buoyancy fields (see line contours in figure \ref{fig:mean_flow}$(a)$ and line plot in figure \ref{fig:mean_flow}$(c)$). Consequently, the mean flow is nearly symmetric about $y=0$ (see figures \ref{fig:mean_flow}$(a)$, \ref{fig:Ro_Ri}$(a)$), with a local Rossby number $Ro_\ell \, (=-\partial_y U) \ll 1$. Additionally, weak frontogenesis results in weaker stratification and vertical shear, yielding a local Richardson number $Ri_\ell$ very close to $1$. For large $Ro$ values, geostrophic adjustment drives much stronger frontogenesis which leads to stronger horizontal buoyancy gradients near the top and bottom surfaces of the domain (line contours in figures \ref{fig:mean_flow}$(b,d)$). The corresponding along-front mean flow is asymmetric about $y=0$ (figure \ref{fig:mean_flow}$(b)$), producing a pronounced asymmetry in the vorticity (figure \ref{fig:Ro_Ri}$(b)$). In this regime, the frontal region is characterized by $|Ro_\ell| \sim \mathcal{O}(1)$ and $Ri_\ell \lesssim \mathcal{O}(1)$ (figures \ref{fig:Ro_Ri}$(b,d)$) \--  characteristic of submesoscale oceanic fronts.

The necessary condition for instability of a baroclinic flow is a sign change in the isopycnal gradient of PV within the domain \citep{eliassen1983charney}. The adjusted front solution described by (\ref{ou1984_sol}$a-c$) satisfies 
this condition for all $Ro$ values considered (not shown), indicating that the basic state is susceptible to BCI.

%%%%%%
%%%%%%
%%%%%%%
%%%%%%%
% \begin{figure}
%     \centering
%     \includegraphics[width=\linewidth]{isopync_PVgrad.png}
%     \caption{Along-isopycnal PV gradient $\partial_s Q$ given by (\ref{isopyc_Q}) for $(a)$ $\beta=0.1$ and $(b)$ $\beta=2$. For the case of $\beta=0.1$, $\partial_s Q$ changes sign across $y=0$ and is independent of $z$, whereas for $\beta=2$, $\partial_s Q$ is confined to the cyclonic vorticity regions at the top and bottom of the domain (e.g., figure \ref{fig:Ro_Ri}$(c)$).}
%     \label{fig:isopync_PV}
% \end{figure}

\subsection{Linear stability analysis}
\label{perturb}
The solution described by (\ref{ou1984_sol}$a-c$) breaks down when isopycnals intersect, i.e., when $\partial_y \eta =0$. From (\ref{ou1984_sol}$(a)$), (\ref{buoy_front}), this occurs at the lower and upper surfaces of the domain when $|\partial_\eta ^2 B|=2$. We can therefore obtain  non-singular solutions to (\ref{ou1984_sol}$a-c$) when $\beta \lesssim 2.28$ (i.e., $Ro \lesssim 1.3$) and perform a linear stability analysis in the range $2.5\times10^{-3} \leq Ro \leq 1$. To this end, we consider the evolution of infinitesimal perturbations to the geostrophically adjusted frontal flow described in \S \ref{mean_flow}. The resulting nondimensional, linearized Boussinesq equations of motion under the $f$-plane approximation are given by
\begin{subequations}
\label{lin_eq}
\begin{align}
\label{mom_eq}
    \frac{D \mathbf{u}}{Dt}
    + \Big(v \frac{\partial U}{\partial y} + w \frac{\partial U}{\partial z} \Big) \hat{x}
    + \hat{z} \times \mathbf{u} &=
    -\nabla p + \frac{1}{\epsilon} b \hat{z} + E \nabla^2 \mathbf{u}, 
\\
\label{buoy_eq}
    % \frac{\partial b}{\partial t} + U \frac{\partial b}{\partial x}  
    \frac{Db}{Dt}
    +  v \frac{\partial B}{\partial y} + w \frac{\partial B}{\partial z} &= \frac{E}{Pr} \nabla^2 b,
\\
\label{contiuity_eq}
    \nabla \cdot \mathbf{u} &= 0,
\end{align}
\end{subequations}
where $D/Dt \equiv \partial/\partial t + U (\partial/\partial x)$ is the material derivative, $\mathbf{u} \equiv (u, v, \epsilon w)$ is the velocity perturbation, $\epsilon=H/R$ is the aspect ratio, $p$ is the pressure perturbation, and $b$ is the buoyancy perturbation. The gradient and Laplacian operators are $\nabla \equiv (\partial/\partial x, \partial/\partial y, (1/\epsilon) \partial/\partial z)$, and $\nabla^2 \equiv \partial^2/\partial x^2 + \partial^2/\partial y^2 + (1/\epsilon^2) \partial^2/ \partial z^2$. In what follows, we compare between hydrostatic and non-hydrostatic flow regimes by choosing $\epsilon=0.1$ and $\epsilon=1$, respectively.
The Ekman number $E=\nu/fR^2$ is set to be $10^{-8}$, and the Prandtl number $Pr=\nu/\kappa$ is taken to be $1$. 
To eliminate pressure, following \cite{teed2010rapidly}, we apply the operator $\hat{z} \cdot \nabla \times \nabla \times $  and $\hat{z} \cdot \nabla \times$ to the momentum equation (\ref{mom_eq}). This procedure yields governing equations of three perturbation variables, the vertical velocity $w$, the vertical vorticity $\zeta \, (=\hat{z} \cdot \nabla \times \mathbf{u})$, and the buoyancy $b$, 
\begin{subequations}
\label{govern_eqs}
\begin{align}
    \frac{D}{Dt}
    \nabla^2 {w} + 
    \Big(\frac{\partial^2 U}{\partial y^2} 
    -\frac{1}{\epsilon^2} \frac{\partial^2 U}{\partial z^2} \Big) \frac{\partial w}{\partial x}
    + 2 \frac{\partial U}{\partial y}
    \Big(\frac{\partial^2 w}{\partial x \partial y} - \frac{1}{\epsilon^2} \frac{\partial^2 v}{\partial x \partial z} \Big)
    \nonumber \\
    -{\frac{2}{\epsilon^2} \frac{\partial^2 U}{\partial y \partial z} \frac{\partial v}{\partial x}} 
    + \frac{1}{\epsilon^2} \frac{\partial \zeta}{\partial z} 
    = \frac{1}{\epsilon^2} \nabla_h^2 b + E \nabla^4 w,
\\
    \frac{D \zeta}{Dt}
    + \frac{\partial U}{\partial y} \frac{\partial w}{\partial z} 
    - \frac{\partial U}{\partial z}
    \frac{\partial w}{\partial y}
    - \frac{\partial^2 U}{\partial y \partial z} w
    -\frac{\partial^2 U}{\partial y^2} v  
    - \frac{\partial w}{\partial z} = 
    E \nabla^2 \zeta, 
\\
    \frac{Db}{Dt}
    + v \frac{\partial B}{\partial y} + 
    w \frac{\partial B}{\partial z}
    = \frac{E}{Pr} \nabla^2 b,
\end{align}
\end{subequations}
where $\nabla_h^2 \equiv \partial^2 /\partial x^2 + \partial^2/\partial y^2$. The benefit of using (\ref{govern_eqs}$(a-c)$) over (\ref{lin_eq}$(a-c)$) is that it enables us to examine the instability at an along-front wavenumber $k \to 0$. The horizontal velocities $u$ and $v$ are related to the vertical velocity $w$ and vertical vorticity $\zeta$ by the identities, 
\begin{subequations}
\label{uv_relation}
\begin{align}
    \nabla_h^2 u &= -\frac{\partial \zeta}{\partial y} - \frac{\partial^2 w}{\partial x \partial z}, 
\\
    \nabla_h^2 v &= \frac{\partial \zeta}{\partial x} - \frac{\partial^2 w}{\partial y \partial z}.    
\end{align}
\end{subequations}
In deriving the above equations, we make use of the continuity equation (\ref{contiuity_eq}) and the definition of vertical vorticity $\zeta$.

To facilitate comparison, we also perform QG stability analysis based on the same underlying frontal flow. The details of the QG stability setup are provided in Appendix \ref{qg_stability}. In addition, we compare the stability results with the classical solutions of \citet{eady1949long} and S71.
%\blue{focusing on the broad-front regime (i.e., lower values of $Ro$). ISN'T RO IDENTICALLY ZERO IN THESE CONFIGURATIONS?}.

\subsubsection{Normal mode equations}
Next, we consider normal-mode perturbations of the form 
\begin{align}
\label{normal_mode}
    [w, \, \zeta, \, b](x,y,z,t) = \mathfrak{R}\big([\widetilde{w}, \, \widetilde{\zeta}, \, \widetilde{b}](y, z) \, \ee^{\ii kx + \sigma t}\big),
\end{align}
where the symbol $\mathfrak{R}$ denotes the real part and a variable with `tilde' denotes an eigenfunction. The variable $\sigma=\sigma_r + \ii \sigma_i$, where the real part $\sigma_r$,  
represents the growth rate, and the imaginary part $\sigma_i$, 
represents the frequency of the unstable perturbation.

After introducing the form  (\ref{normal_mode})  into the governing equations (\ref{govern_eqs}$a-c$), the following system of differential equations is obtained.
\begin{subequations}
\label{eigen_form}
\begin{align}
\label{eigen_form_a}
    \left[(\ii k U - E \mathcal{D}^2) \mathcal{D}^2 \widetilde{w} + \ii k \left(\partial^2_y {U} - \epsilon^{-2} \partial^2_z {U} \right) \widetilde{w} + 2 \ii k \partial_y U \partial_y \widetilde{w}
     \right] + \epsilon^{-2} \partial_z \widetilde{\zeta}
     \nonumber \\
     - (2 \ii k \epsilon^{-2})  \partial_y U 
     \partial_z \widetilde{v}
     - {(2 \ii k \epsilon^{-2}) \partial_{yz} U \widetilde{v}} 
     - \epsilon^{-2} \mathcal{D}_h^2 \widetilde{b}
    = -\sigma \mathcal{D}^2 \widetilde{w},
\\
\label{eigen_form_b}
    \left[-\partial_{yz} U \widetilde{w} - \partial_z U \partial_y \widetilde{w} + (\partial_y U -1)\partial_z \widetilde{w} \right] +
    \left[ik U - E \mathcal{D}^2 \right] \widetilde{\zeta}
    -\partial^2_{y} U \widetilde{v} = -\sigma \widetilde{\zeta},
\\
\label{eigen_form_c}
    \partial_z B \widetilde{w} + \partial_y B  \widetilde{v} + 
    \left[ik U - E \mathcal{D}^2 \right] \widetilde{b} = -\sigma \widetilde{b}, 
\end{align}
\end{subequations}
where $\mathcal{D}^4  = (\mathcal{D}^2 )^2 =\big(\partial_y^2 + (1/\epsilon^2)\partial_z^2 - k^2\big)^2$ and $\mathcal{D}_h^2 = (\partial_y^2 - k^2)$.
The eigenfunctions $\widetilde{u}$, $\widetilde{v}$ are related to  $\widetilde{w}$, $\widetilde{\zeta}$ by the relations from (\ref{uv_relation}$(a,b)$), 
\begin{subequations}
\label{identity}
\begin{align}
    -\mathcal{D}_h^2 \widetilde{u} &= \ii k \partial_{z} \widetilde{w} + \partial_y \widetilde{\zeta},
\\   
    -\mathcal{D}_h^2 \widetilde{v} &= \partial_{yz} \widetilde{w} -  \ii k \widetilde{\zeta}.
\end{align}
\end{subequations}
We apply periodic boundary conditions in the $y$ direction and  free-slip, rigid lid, and zero buoyancy gradient boundary conditions in the $z$ direction, i.e., 
\begin{align}
\label{bcs}
    \widetilde{w} = \partial_{zz} \widetilde{w} = 
    \partial_z \widetilde{\zeta} = \partial_z \widetilde{b} = 0, 
    \,\,\,\,\,\,\, \text{at} \,\,\, {z}=0, 1.
\end{align}
Equations (\ref{eigen_form}$a-c$), with (\ref{identity}$a-b$) and (\ref{bcs}) can be expressed as a standard generalized eigenvalue problem,
\begin{align}
\label{gen_eigvals}
    \bm{\mathsfit{A}} \bm{\mathsf{X}}=
    {\sigma} \bm{\mathsfit{B}} \bm{\mathsf{X}},   
\end{align}
where $\sigma$ is the eigenvalue, $\bm{\mathsf{X}}=[\widetilde{w}, \widetilde{\zeta}, \widetilde{b}]^T$ is the eigenvector and the matrices $\bm{\mathsfit{A}}$, $\bm{\mathsfit{B}}$ are the complex and real non-symmetric matrices, respectively. The elements of matrices $\bm{\mathsfit{A}}$ and $\bm{\mathsfit{B}}$ are shown in Appendix \ref{matrices_details}. We solve the above eigenvalue problem following the procedure discussed in the next section.

\subsubsection{Numerical method}
\label{method}
To solve the eigenvalue problem (\ref{gen_eigvals}), a spectral collocation method is used that utilizes Chebyshev differentiation in the $z$ direction and Fourier differentiation in the $y$ direction \citep{trefethen2000spectral}. The generalized eigenvalue problem in (\ref{gen_eigvals}) is solved using the FEAST algorithm, which is based on the complex contour integration method \citep{polizzi2009density}. The benchmark of the eigensolver is presented in Appendix \ref{benchmark}.

To minimize the influence of periodic boundary conditions on the stability solution, the cross-front domain length is set to $3Ro^{-1/2}$, which ensures sufficient domain size (e.g., figure \ref{fig:eigfuns_cmp}). The grid independent tests of the stability results are discussed in Appendix \ref{grid_test}. Unless otherwise stated, all results shown hereafter use $N_y = 240$ and $N_z = 32$, where $N_y$ and $N_z$ denote the number of points in the $y$ and $z$ directions, respectively.

\subsubsection{Kinetic energy equation}
\label{pe_ke}
The governing equation of the perturbation KE is given by
\begin{align}
\label{pertb_ke}
    2 \sigma \left \langle K \right \rangle_x
    + \underbrace{\left \langle \widetilde{u} \widetilde{v}^\star - \widetilde{u}^\star \widetilde{v} \right \rangle_x}_{{\text{Coriolis}}} 
    = \underbrace{-\left \langle \widetilde{u}^\star \widetilde{v} \frac{\partial {U}}{\partial {y}} \right \rangle_x}_{{{\text{HSP}}}}
     \underbrace{-\left \langle \widetilde{u}^\star \widetilde{w} \frac{\partial {U}}{\partial {z}} \right \rangle_x}_{{{\text{VSP}}}}
     + \underbrace{\left \langle \widetilde{w}^\star \widetilde{b} \right \rangle_x}_{{{\text{BFLUX}}}}
     + \underbrace{\left \langle \widetilde{\nabla} \cdot  \left(\widetilde{\bm{u}}^\star \widetilde{p} \right) \right \rangle_x}_{{\text{PWORK}}}
     \nonumber \\
     + \underbrace{\left \langle E \left(\widetilde{u}^\star \nabla^2 \widetilde{u} 
     + \widetilde{v}^\star {\nabla}^2 \widetilde{v} 
     + \epsilon^2 \widetilde{w}^\star {\nabla}^2 \widetilde{w} \right) \right \rangle_x}_{{\text{DISP}}},
\end{align}
where $\langle \cdot \rangle_x$ denotes the $x$ integral over one wavelength. The perturbation KE $K$ is defined as $K=(\widetilde{u} \widetilde{u}^\star + \widetilde{v} \widetilde{v}^\star + \epsilon^2 \widetilde{w} \widetilde{w}^\star)/2$, with the superscript `star' denoting a complex conjugate quantity. 
The term {Coriolis} in (\ref{pertb_ke}) is purely imaginary and thus does not contribute to the growth of the perturbation KE. 
The first two terms on the right-hand side of (\ref{pertb_ke}), horizontal shear production (HSP) and vertical shear production (VSP), are associated with the horizontal and vertical shear of the mean flow, respectively. A positive value of HSP (or VSP) describes the growth of the perturbation KE at the expense of the mean flow KE. The third term on the right-hand side of (\ref{pertb_ke}), the buoyancy flux (BFLUX), quantifies energy exchanges between the perturbation kinetic and potential energies. The pressure work (PWORK) term denotes the propagation of KE due to pressure perturbations and has a zero domain average. The dissipation term (DISP) for the unstable perturbation is negligible due to the small value of $E$ in the stability analysis (not shown). 

Throughout the manuscript, we define the domain integral (in $y$-$z$ plane) of a quantity $\phi$ as
\begin{align}
\label{domain_integ}
    \langle \phi \rangle_{yz} = \int_{-1.5L_f}^{1.5L_f} \int_0^1 \phi \, dy dz, 
\end{align} 
and denote the volume integral of the quantity $\phi$ by $\langle \phi \rangle_{xyz}$.

\begin{figure}
    \centering
    \includegraphics[width=\linewidth]{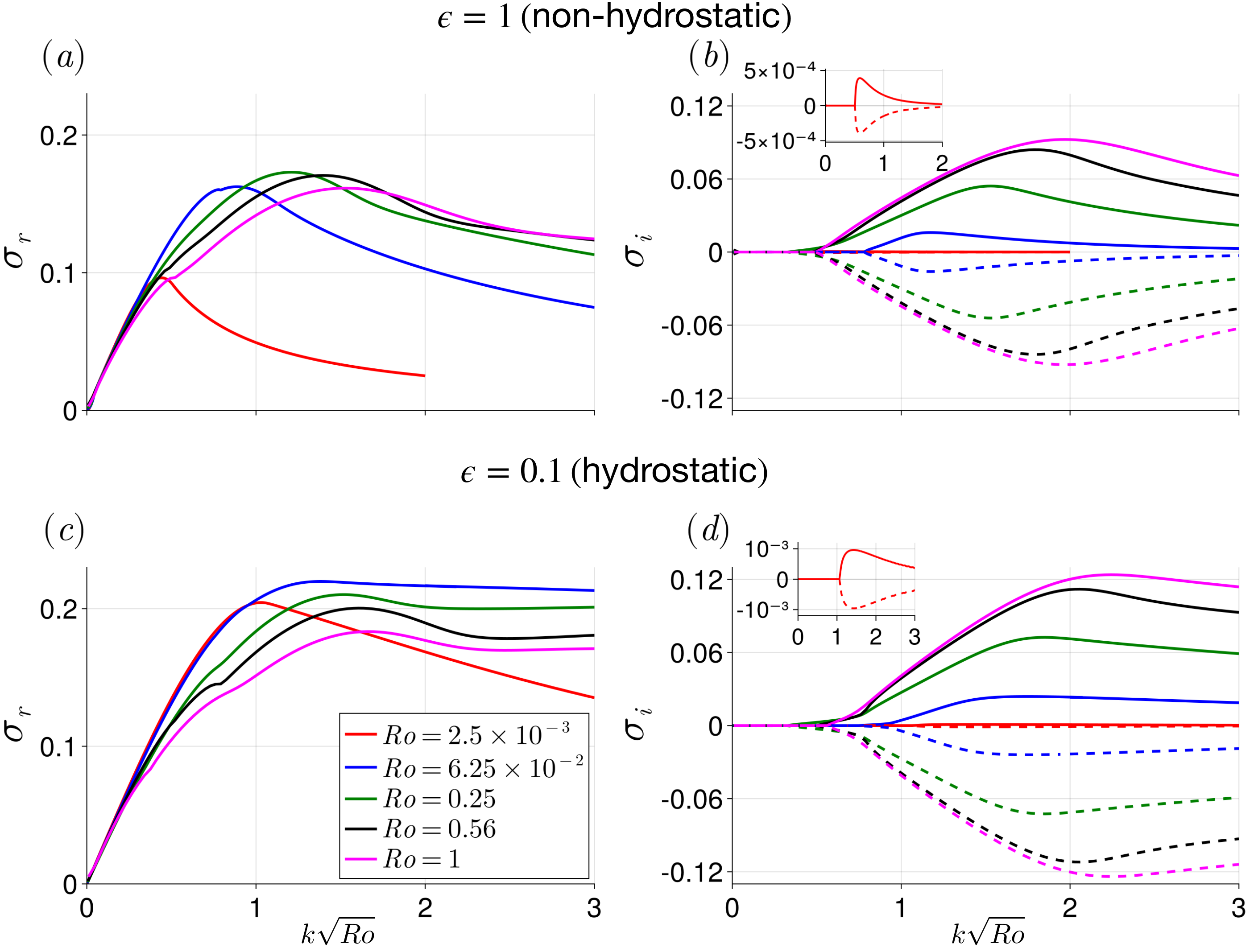}
    \caption{The nondimensional growth rate $\sigma_r$ (panels $(a,c)$) and corresponding frequency $\sigma_i$ (panels $(b,d)$) for different $Ro$ values with $\epsilon=1$ (top panels; non-hydrostatic regime) and $\epsilon=0.1$ (bottom panels; hydrostatic regime). The insets in panels $(b,d)$ zoom in on the frequency diagrams for the case of $Ro=2.5\times10^{-3}$. Note that the wavenumber $k$ is non-dimensionalized by the dimensional cross-frontal width ${Ro}^{-1/2} R$, the natural length scale of the adjusted front.}
    \label{fig:growthrate_k}
\end{figure}
%%%%
%%%%
\begin{figure}
    \centering
    \includegraphics[width=\linewidth]{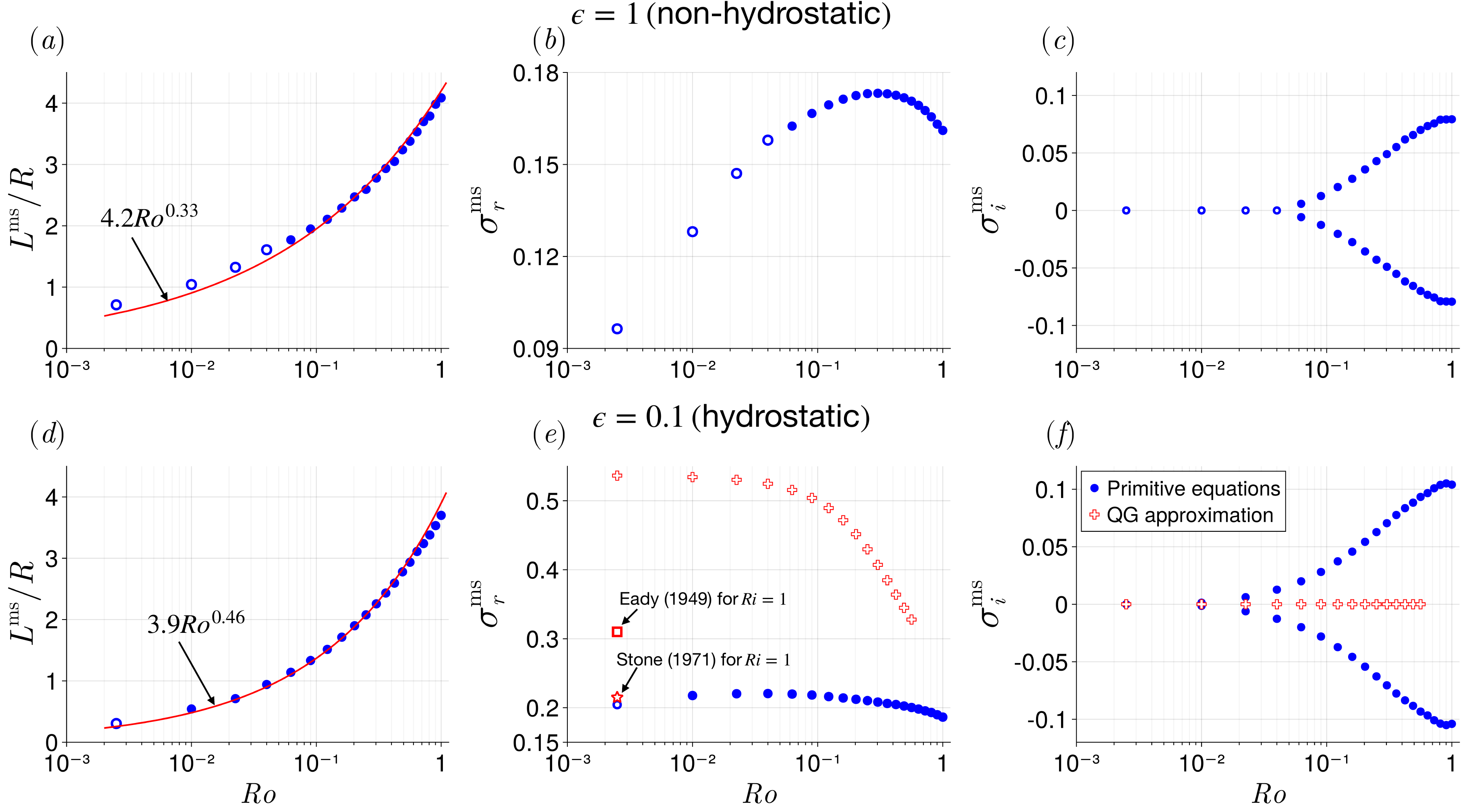}
    \caption{{Panel $(a,d)$ shows the non-dimensional wavelength of the most unstable mode $L^\text{ms}/R=(2\pi/k^\text{ms})$}. The red lines in panels $(a,d)$ represents the least-squares fits of $L^\text{ms}/R$ as a function of $Ro$. 
    The nondimensional growth rate of the most unstable mode, $\sigma_r^\text{ms}$ (panels $(b,e)$), and the corresponding frequency, $\sigma_i^\text{ms}$ (panels $(c,f)$), are shown for various $Ro$ values. The top (bottom) panel shows for the non-hydrostatic (hydrostatic) regime. Open blue circles denote BCI modes, while filled circles represent ICL instability modes. Open crosses in panels $(c)$ and $(d)$ indicate growth rates and frequencies obtained from the QG stability analysis (Appendix \ref{qg_stability}). For comparison, the growth rate of most unstable BCI mode of  the S71 (red star) and Eady (red rectangle) solutions are shown for $Ri=1$ and $Ro=0$ (figure \ref{fig:benchmark}).  
    % Open red star in panel $(c)$ marks the theoretical growth rate of the most unstable mode for $Ri=1$, based on the analytical solution of \citet{S71baroclinic} (). Open red rectangle in panel $(c)$ highlights the most unstable growth rate predicted by the Eady solution \citep[see Figure \ref{fig:benchmark}$(a)$ of Appendix \ref{benchmark}]{eady1949long}. 
    The $x$-axes in all panels are shown on a logarithmic scale.
    % The right axis of panel $(a)$ displays the Rossby number ($Ro_\text{bulk}$ shown by magenta diamonds and $Ro_\text{front}$ by magenta triangles). The right axis of panel $(b)$ displays the Richardson number ($Ri_\text{bulk}$ indicated by magenta diamonds and $Ri_\text{front}$ by magenta triangles). The `front' values are based on the rms value calculated over the \red{upper (so it the 90th percentile only the top half of the domain? why do you only consider the top half? )} frontal region, defined using the $90$th percentile of horizontal buoyancy gradient magnitudes, $|\partial_y B|$. The `bulk' values are rms estimates computed over the domain bounded by $y \in [-L_f, L_f]$ and $z \in [0, 1]$. 
    }
    \label{fig:growth_rate}
\end{figure}
%%%%%
%%%%%
%%%%%%%%%%%%%%%%%%
%%%%% table
%%%%%%%%%%%%%%%%%%
\begin{table}%[!htp]
\centering
\caption{The complex frequency $\sigma^\text{ms}$ and corresponding wavenumber $k^\text{ms}$ of the most unstable mode for different values of $Ro$. Note that  $k^\text{ms}$ is {non-dimensionalized} by the cross-frontal width ${Ro}^{-1/2} R$. Results are presented for two regimes, $\epsilon=1$ and $\epsilon=0.1$. Results from the QG linear stability analyses are also shown for four values of $Ro$. The zero frequency modes are the BCI modes. }
\label{tab:growthrate_k_tab}
\scriptsize
\vspace{0.3mm}
\begin{tabular}{crrrrrrrr}
\toprule
\multirow{2}{*}{\small $Ro$}
&\multirow{2}{*}{\small $(Ro_\ell)_\text{max}$}
&\multicolumn{2}{c}{\small $\epsilon=1$} 
&\multicolumn{2}{c}{\small $\epsilon=0.1$} 
&\multicolumn{2}{c}{\small QG} 
\\\cmidrule{3-8}
& & {\small $k^\text{ms}\sqrt{Ro}$} 
& \multicolumn{1}{c}{\small $\sigma^\text{ms}$} 
& {\small $k^\text{ms}\sqrt{Ro}$} 
& \multicolumn{1}{c}{\small $\sigma^\text{ms}$} 
& {\small $k^\text{ms}\sqrt{Ro}$} 
& \multicolumn{1}{c}{\small $\sigma^\text{ms}$} 
\\\midrule
\multicolumn{1}{c}{\small $2.5\times10^{-3}$} 
& \multicolumn{1}{c}{\small $2\times10^{-3}$} 
& \multicolumn{1}{c}{\small $0.441$} 
& \multicolumn{1}{c}{\small $0.096$} 
& \multicolumn{1}{c}{\small $1.031$} 
& \multicolumn{1}{c}{\small $0.204$} 
& \multicolumn{1}{c}{\small $2.783$} 
& \multicolumn{1}{c}{\small $0.537$} 
\\
\multicolumn{1}{c}{\small $6.25\times10^{-2}$} 
& \multicolumn{1}{c}{\small $0.05$} 
& \multicolumn{1}{c}{\small $0.887$} 
& \multicolumn{1}{c}{\small $0.163 \pm 0.006 \ii$} 
& \multicolumn{1}{c}{\small $1.378$} 
& \multicolumn{1}{c}{\small $0.219 \pm 0.021 \ii$} 
& \multicolumn{1}{c}{\small $2.753$} 
& \multicolumn{1}{c}{\small $0.525$} 
\\
\multicolumn{1}{c}{\small $0.25$} 
& \multicolumn{1}{c}{\small $0.25$} 
& \multicolumn{1}{c}{\small $1.211$} 
& \multicolumn{1}{c}{\small $0.173 \pm 0.043 \ii$} 
& \multicolumn{1}{c}{\small $1.513$} 
& \multicolumn{1}{c}{\small $0.210 \pm 0.063 \ii$} 
& \multicolumn{1}{c}{\small $2.556$} 
& \multicolumn{1}{c}{\small $0.429$} 
\\
\multicolumn{1}{c}{\small $0.56$} 
& \multicolumn{1}{c}{\small $0.76$} 
& \multicolumn{1}{c}{\small $1.394$} 
& \multicolumn{1}{c}{\small $0.171 \pm 0.071 \ii$} 
& \multicolumn{1}{c}{\small $1.606$} 
& \multicolumn{1}{c}{\small $0.201 \pm 0.097 \ii$} 
& \multicolumn{1}{c}{\small $2.347$} 
& \multicolumn{1}{c}{\small $0.328$} 
\\
\multicolumn{1}{c}{\small $1$} 
& \multicolumn{1}{c}{\small $3.23$} 
& \multicolumn{1}{c}{\small $1.538$} 
& \multicolumn{1}{c}{\small $0.161 \pm 0.079\ii$} 
& \multicolumn{1}{c}{\small $1.698$}
& \multicolumn{1}{c}{\small $0.188 \pm 0.106\ii$} 
& \multicolumn{1}{c}{\small $-$} 
& \multicolumn{1}{c}{\small $-$}
\\
\bottomrule
\end{tabular}
\end{table}
%%%%%%%
%%%%%%%
%%%%%%%
\section{Results}
\label{results}
We examine the linear stability analyses of the basic state described by (\ref{ou1984_sol}$a-c$) across a range of Rossby numbers (by varying $\beta \in  [0.1, 2]$ with increments of $0.1$), for two values of $\epsilon$, $\epsilon=1$ (non-hydrostatic) and $\epsilon=0.1$ (hydrostatic). The nondimensional growth rate $\sigma_r$ and the corresponding frequency $\sigma_i$ as a function of the along-front wavenumber $k$ for five different values of $Ro$ are shown  in figure \ref{fig:growthrate_k}.
In contrast to the QG BCI described by \citet{eady1949long} and the non-geostrophic BCI analysis of \citet{stone1966non}, the present analysis exhibits no short-wave cut-off in the perturbation growth rate (figures \ref{fig:growthrate_k}$(a,c)$). 
Notably, the frequency–wavenumber diagrams (figures \ref{fig:growthrate_k}$(b,d)$) reveal two distinct regimes \-- zero frequency and non-zero frequency (propagating modes).

The zero frequency regime corresponds to BCI, which can be interpreted as the phase locking of two counter-propagating Rossby waves situated in regions of opposing isopycnal PV gradients \citep{hoskins1985use}. In contrast, the non-zero frequency mode is identified as the inertia-critical layer (ICL) instability mode (see \S \ref{eigenmode_cmp} for mode structure and discussion). The ICL instability mechanism is associated with a singularity at an ICL, where the Doppler-shifted frequency matches plus or minus the Coriolis frequency \citep{jones1967propagation}. 
In nondimensional form, this condition is expressed as 
\begin{align}
\label{def_icl}
    k(U \pm \Delta U) - \sigma_i = \pm 1,  
\end{align}
where $\Delta U = U_\text{max} - U_\text{min}$, with $U_\text{max}$ and $U_\text{min}$ denoting the global maximum and minimum values of $U$, respectively (e.g., figure \ref{fig:flux}$(a)$). The positive (negative) sign corresponds to perturbation frequency with positive (negative) $\sigma_i$ in (\ref{def_icl}). Physically, the ICL acts as an absorber of inertia-gravity waves, facilitating the transfer of wave momentum to the mean flow \citep{jones1967propagation}. The positive and negative frequency ICL instability modes have identical growth rates (figures \ref{fig:growthrate_k}$(a,c)$), with a positive (negative) frequency corresponding to a phase speed in the positive (negative) $x$-direction. 

In the non-hydrostatic case, the growth rate, $\sigma_r$, decreases for large values of $k$ for all $Ro$ values (figure \ref{fig:growthrate_k}$(a)$). In contrast, in the hydrostatic case $\sigma_r$ remains relatively constant for large $k$ values (with the exception of $Ro=2.5\times10^{-3}$) indicating a broad range of unstable modes with comparable growth rates (figure \ref{fig:growthrate_k}$(c)$).

For the basic state considered by S71 the front has infinite width, whereas in the present analysis the cross-front width scales as ${Ro}^{-1/2}$. Consequently, at lower $Ro$ value (i.e., $Ro = 2.5\times10^{-3}$, $Ri {=} 1$; figure \ref{fig:Ro_Ri}$(b)$), the growth rate of the most unstable BCI mode, {$\sigma_r^\text{ms}$}, approaches that reported by S71 (open blue circle for $Ro=2.5\times10^{-3}$ and open red star in figure \ref{fig:growth_rate}$(c)$), while the Eady solution has a larger growth rate (open red rectangle in figure \ref{fig:growth_rate}$(c)$), likely due to the fact that the unstable mode is ageostrophic when $Ri=1$. 
%Notably, the inclusion of horizontal shear in the present stability analysis leads to a substantial modification of the perturbation structure in comparison to the S71 solution (see \S \ref{eigenmode_cmp}).
%%%
In contrast to the hydrostatic case, the most unstable mode in the non-hydrostatic regime remains of BCI type up to $Ro = 0.04$, with {$\sigma_r^\text{ms}$} increasing as $Ro$ increases (figure \ref{fig:growth_rate}$(a)$). This is a consequence of more APE in the frontal region at larger $Ro$ (not shown). Non-hydrostatic effects also act to reduce the most unstable BCI wavenumber, $k^\text{ms}$; for instance, at $Ro=2.5\times10^{-3}$, $k^\text{ms}$ decreases from approximately $20.61$ to $8.83$ as $\epsilon$ increases from $0.1$ to $1$ 
(see table \ref{tab:growthrate_k_tab} for comparison at $Ro=2.5\times10^{-3}$).

As $Ro$ increases, the dominant instability transitions to the ICL instability mode. 
%For smaller values of $Ro$, the ICL instability mode appears only at higher $k$, whereas for larger $Ro$, it emerges at $k \sim \mathcal{O}(1)$ (figure \ref{fig:growthrate_k}$(b,d)$ and table \ref{tab:growthrate_k_tab}). 
The growth rate of the most unstable ICL mode is slightly higher in the hydrostatic case compared to the non-hydrostatic case (table \ref{tab:growthrate_k_tab}). The growth rate of the most unstable ICL instability mode is largely independent of $Ro$ in both hydrostatic and non-hydrostatic cases, while the wavenumber corresponding to the most unstable mode decreases with $Ro$ in both cases. The wavenumber of the most unstable mode is consistently slightly larger in the hydrostatic case than in the non-hydrostatic case (table \ref{tab:growthrate_k_tab}).

The QG stability analysis consistently overpredicts  $\sigma_r^\text{ms}$ (open plus symbols in figure \ref{fig:growth_rate}$(c)$). The higher growth rate of the QG BCI mode at the low $Ro$ is because the adjusted front exhibits strong horizontal variations in the stratification, while the QG analysis assumes that changes in the stratification are small.\footnote{
{as expected, the low Rosssby number QG BCI growth rate approaches that of Eady when using the same stratification value.}}
%For example when $Ro=2.5\times10^{-3}$ the maximum magnitude of $Ro^{-1/2}|\partial_{yz} B|$ is an order of magnitude larger than the maximum magnitude of $\partial_z B$).
The growth rate of the most unstable QG BCI mode decreases monotonically with increasing $Ro$, reflecting a suppression of BCI by horizontal shear. This is a well-documented effect, commonly referred to as the barotropic governor mechanism \citep{james1986concerning,james1987suppression}. Interestingly, when the most unstable QG BCI wavenumber is non-dimensionalized by the frontal width ${Ro}^{-1/2} R$, it remains largely insensitive to $Ro$ (table \ref{tab:growthrate_k_tab}). This implies that the cross-front length scale sets the wavelength of the most unstable QG BCI mode. 

\subsection{Mode structure}
\label{eigenmode_cmp}
\begin{figure}
    \centering
    \includegraphics[width=0.94\linewidth]{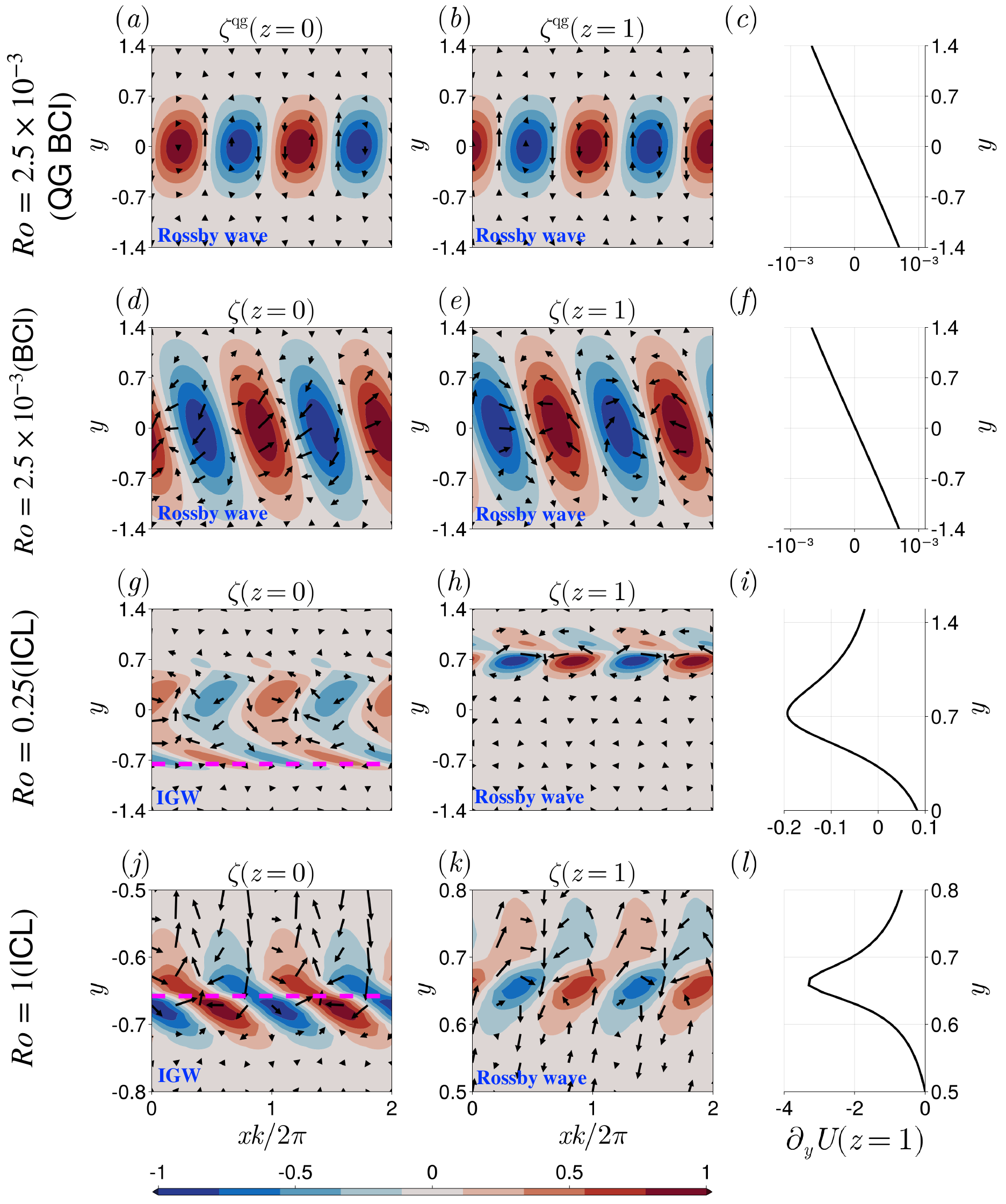}
    \caption{Comparison of perturbation vertical vorticity for the most unstable modes across different $Ro$ values, computed based on the hydrostatic ($\epsilon=0.1$) bi-global stability analysis. Panels $(a,b)$ and $(d,e)$ correspond to the BCI modes computed separately from the QG stability analysis (Appendix \ref{qg_stability}) and from the primitive equation stability analysis, respectively, at $Ro=2.5\times10^{-3}$. Panels $(g,h)$ and $(j,k)$ correspond to the ICL instability modes  for the positive frequency branch, computed at $Ro=0.25$ and $Ro=1$, respectively. The left and middle columns show the vertical vorticity structures in the $x$–$y$ plane at $z=0$ and $z=1$, respectively, while the right column (panels $(c,f,i,l)$) shows the corresponding horizontal shear $\partial_y U$ at $z=1$. The velocity vectors normalized by their respective maximum amplitudes (embedded black arrows) show that the horizontal circulation is anticlockwise (clockwise) around regions of positive (negative) vertical vorticity for the Rossby wave modes. The dotted magenta lines in panels $(g)$ and $(h)$ show the ICL described by (\ref{def_icl}). The $x$-axis in all panels are scaled by the corresponding wavelength $2\pi/k$. Note that the $y$–axis limits in panels $(j,k,l)$ differ from those of all other panels. }
    \label{fig:eigfuns_cmp}
\end{figure}

We can gain insight into the instability and the difference between the primitive equation and QG analysis by considering the spatial structure of the most unstable modes. We begin by analyzing the structure of the most unstable BCI mode with $Ro=2.5\times10^{-3}$ at the bottom ($z=0$) and top ($z=1$) surfaces. Figures \ref{fig:eigfuns_cmp}$(a,b)$ show the most unstable QG BCI mode and figures \ref{fig:eigfuns_cmp}$(d,e)$ show the most unstable primitive equation mode in the hydrostatic case. The vertical relative vorticity is shown in panels (c) and (d) for reference. In both cases, the perturbation modes are confined in the cross-front direction, although the QG solution shows notably stronger cross-front localization than its primitive equations counterpart. The confinement is due to the fact that horizontal shear localizes the isopycnal PV gradient (horizontal gradient for QG dynamics), creating Rossby wave guides that confine the BCI mode \citep[not shown;][]{hoskins1985use}.  
% This cross-front confinement contrasts with the results of S71, wherein the most unstable mode lacks any cross-front length scale (figure \ref{fig:benchmark}$(b)$). 
The BCI mode in the primitive equations exhibits a clear leftward tilt (figures \ref{fig:eigfuns_cmp}$(d,e)$). Conversely, the corresponding QG BCI mode exhibits a weaker, rightward tilt (figures \ref{fig:eigfuns_cmp}$(a,b)$). For $Ro=0.25$, the rightward tilt in the QG BCI mode is more pronounced (figure \ref{fig:eigfun_qg_beta1.5}), demonstrating the effect of horizontal shear on the perturbation mode structure. 
%This tilt produces a counter-gradient horizontal momentum flux, a hallmark of the barotropic governor mechanism.

For the positive-frequency ICL instability mode, the Rossby wave component is centered on the upper boundary of the domain, while the Doppler-shifted IGW is centered on the lower boundary (figures \ref{fig:eigfuns_cmp}$(g,h)$ and $(j,k)$). This vertical arrangement is reversed for the negative-frequency branch (not shown). The IGW mode is confined near the ICL ((\ref{def_icl}); dashed magenta line in figures \ref{fig:eigfuns_cmp}$(g)$ and $(j)$), whereas the Rossby wave is predominantly localized within the cyclonic vorticity region (figures \ref{fig:eigfuns_cmp}$(h,i)$ and $(k,l)$). The Rossby wave mode tilts against the horizontal shear, enabling it to extract energy from the horizontal shear of the frontal flow \-- consistent with the energetic interpretation discussed in the next section.

\begin{figure}
    \centering
    \includegraphics[width=\linewidth]{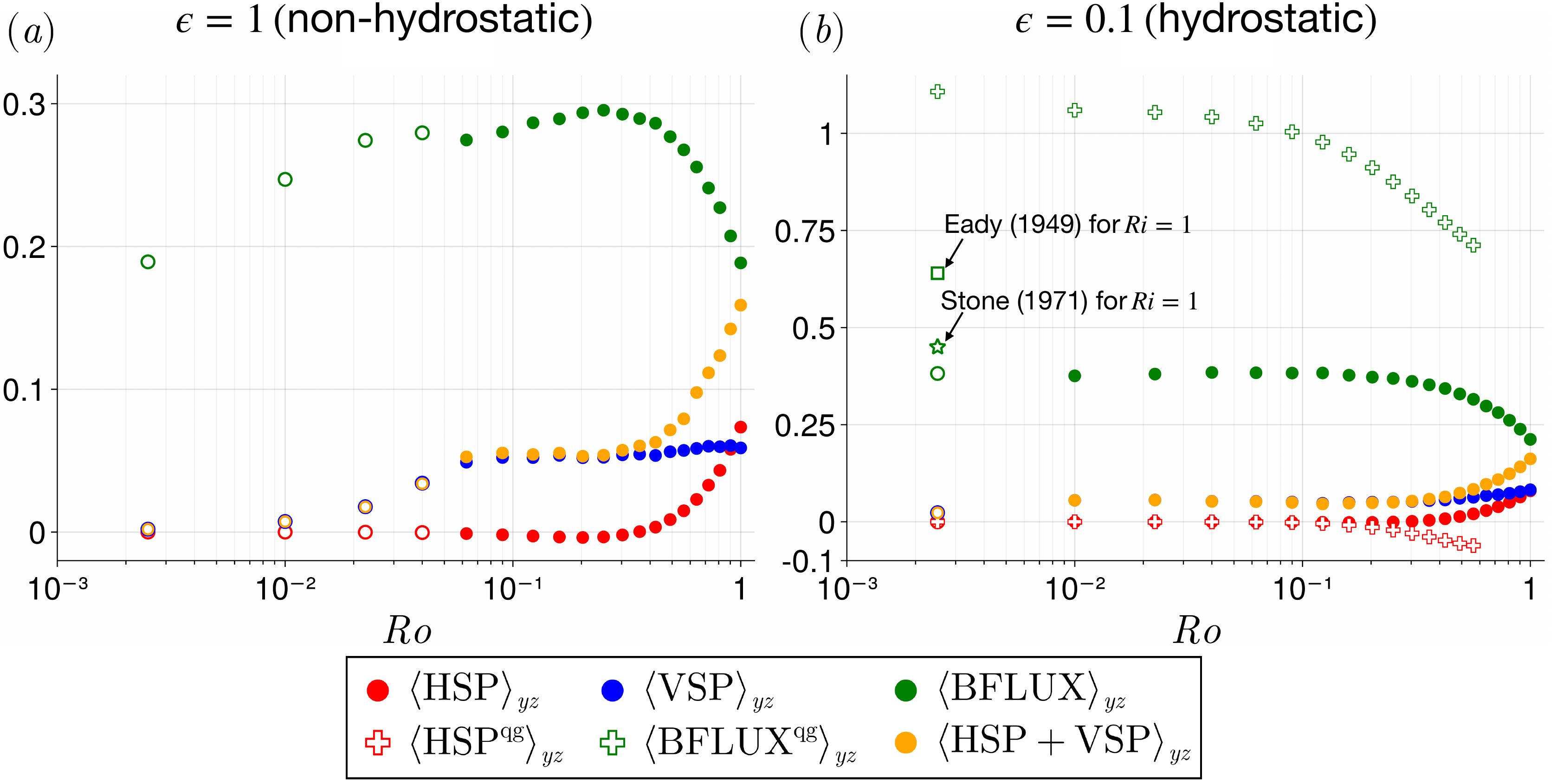}
    \caption{Domain integral (i.e., (\ref{domain_integ})) of the energy exchange terms (\ref{pertb_ke}) for the most unstable mode shown in figure \ref{fig:growth_rate}  for the cases of $(a)$ $\epsilon=1$ and $(b)$ $\epsilon=0.1$. In panels $(a,b)$, open (filled) symbol indicates BCI (ICL) mode. For the ICL modes, the energetics are computed from a linear superposition of the positive and negative frequency modes. For comparison, the corresponding QG energy terms (\ref{qg_ke_eq}) are shown in panel $(b)$ (open red and green crosses). The open green star and rectangle symbols in panel $(b)$ shows the S71 and \citet{eady1949long} solutions, respectively for the case of $Ri=1$ and $Ro=0$.
    The $x$-axes in both panels are shown on a logarithmic scale.}
    \label{fig:energetic}
\end{figure}
%%%%%
%%%%%

\subsection{Energetics}
\label{energetic}
The ICL instability modes consist of conjugate pairs of positive and negative frequency branches, each exhibiting identical growth rates (figure \ref{fig:growth_rate}). From an energetic perspective, it is therefore natural to represent the perturbation solution as a linear superposition of these two modes. To facilitate comparison across different values of $Ro$, all perturbation quantities are scaled such that the volume-integrated perturbation KE equals $1$.

As expected, the primary source of energy driving perturbation growth for the BCI modes is the buoyancy flux (BFLUX), with only minor contributions from the vertical shear production (VSP) and the horizontal shear production (HSP) (figures \ref{fig:energetic}$(a,b)$). For the broad front and low $Ro$ case ($Ro=2.5\times10^{-3}$), $\langle \text{BFLUX} \rangle_{yz}$ closely matches the result of S71 (green circle and green star in figure \ref{fig:energetic}$(b)$). The Eady solution tends to overpredict the buoyancy flux (green rectangle in figure \ref{fig:energetic}$(b)$), again reflecting the fact that the unstable mode is no longer purely geostrophic when $Ri = 1$ \citep{stone1971baroclinic}. In the non-hydrostatic regime, $\langle \text{BFLUX} \rangle_{yz}$ for the BCI mode increases with $Ro$ (open green circles in figure \ref{fig:energetic}$(a)$), consistent with the increase in APE (not shown). Furthermore, for the non-hydrostatic BCI modes ($Ro=2.5\times10^{-3}$ to $0.04$), $\langle \text{VSP} \rangle_{yz}$ remains significantly larger than $\langle \text{HSP} \rangle_{yz}$ (open blue and red circles in figure \ref{fig:energetic}$(a)$; see \S \ref{spatial_structure} for a discussion).

In contrast to the BCI modes, the ICL instability is driven by a combination of $\langle \text{VSP} \rangle_{yz}$, $\langle \text{HSP} \rangle_{yz}$, and $\langle \text{BFLUX} \rangle_{yz}$ (filled circles in figure \ref{fig:energetic}$(a,b)$). For $Ro>0.3$ the contribution from $\langle \text{BFLUX}\rangle_{yz}$ decreases and the contribution from $\langle \text{HSP} \rangle_{yz}$ increases in both hydrostatic and non-hydrostatic cases (filled green circles in figures \ref{fig:energetic}$(a,b)$). In both cases, the sum of the horizontal and vertical shear production becomes comparable to the buoyancy flux when $Ro \sim \mathcal{O}(1)$. 
%In the hydrostatic case, both $\langle \text{HSP} \rangle_{yz}$ and $\langle \text{VSP} \rangle_{yz}$ increase with $Ro$ and eventually become comparable at $Ro \sim \mathcal{O}(1)$ (filled blue and red circles in figure \ref{fig:energetic}$(b)$). In the non-hydrostatic case, by contrast, $\langle \text{VSP} \rangle_{yz}$ remains largely insensitive to $Ro$, while $\langle \text{HSP} \rangle_{yz}$ increases steadily when $Ro > 0.3$, eventually approaching and becoming comparable to $\langle \text{VSP} \rangle_{yz}$ at $Ro \sim \mathcal{O}(1)$ (filled blue and red circles in figure \ref{fig:energetic}$(a)$). 

In the QG solution, perturbation growth is primarily driven by the buoyancy flux ($\langle \text{BFLUX}^\text{qg} \rangle_{yz}$), which decreases monotonically with increasing $Ro$ (green plus in Figure \ref{fig:energetic}$(b)$), an effect attributed to the barotropic governor mechanism. The value of $\langle \text{HSP}^\text{qg} \rangle_{yz}$ is negative (red plus in Figure \ref{fig:energetic}$(b)$), indicating a counter-gradient horizontal momentum flux. Its magnitude increases with $Ro$, consistent with the intensification of horizontal shear.

\begin{figure}
    \centering
    \includegraphics[width=\linewidth]{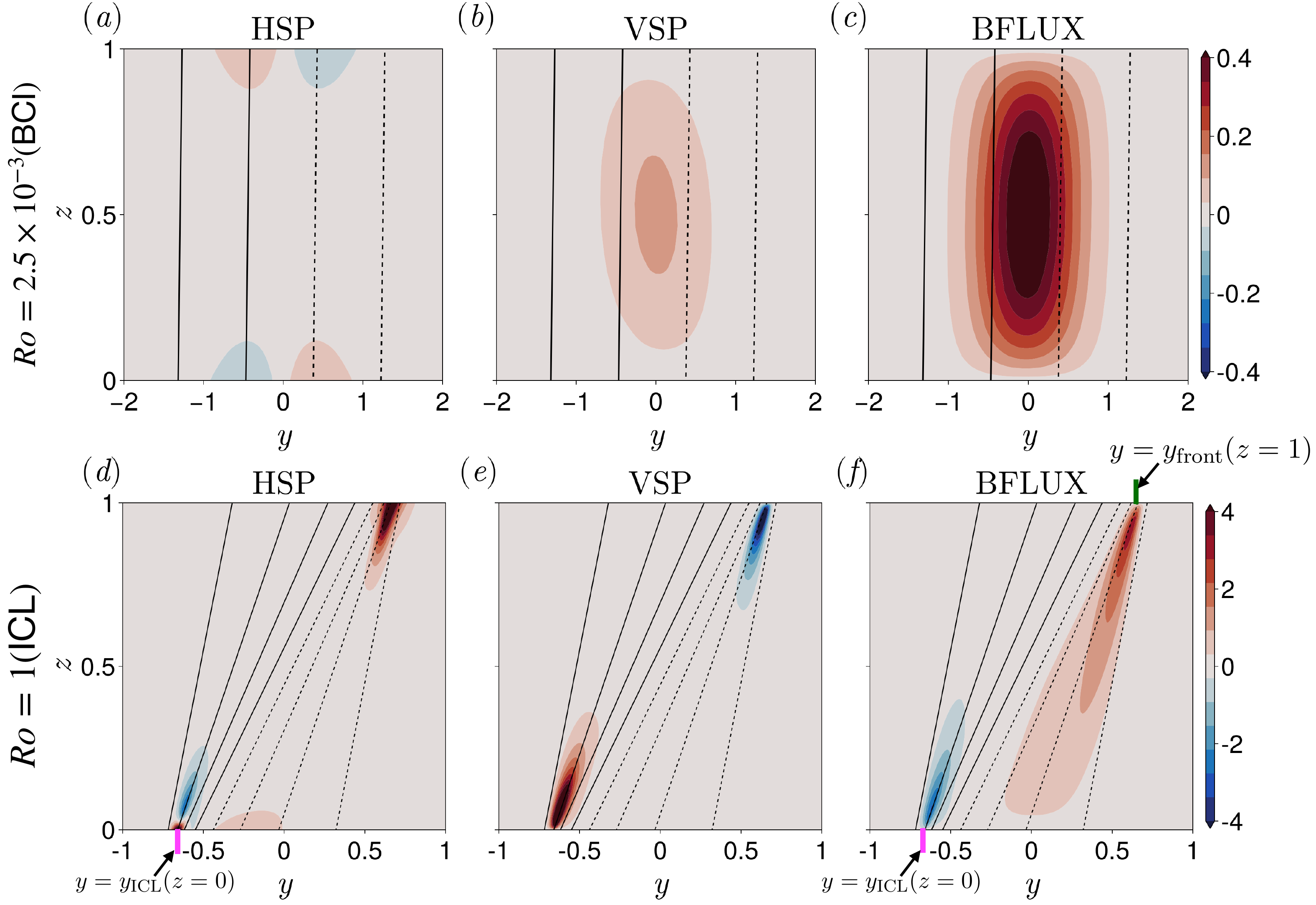}
    \caption{Color contour plots of energy exchange terms in the perturbation KE equation (\ref{pertb_ke}). Panels $(a$–$c)$ show the BCI mode at $Ro=2.5\times10^{-3}$, while panels $(d$–$f)$ correspond to the ICL instability mode at $Ro=1$, both under hydrostatic conditions. The black lines represent the adjusted isopycnals $B$, with contour intervals of $0.04$ in panel $(a$-$c)$ and $0.11$ in panel $(d$-$f)$. The solid (dashed) black lines indicate a positive (negative) values. The vertical magenta lines in panels $(d)$ and $(f)$ denote the location of the ICL \-- $y_\text{ICL} = -0.67$ at $z = 0$. The vertical green line in panel $(c)$ marks the location of the maximum horizontal buoyancy gradient\-- $y_\text{front} = 0.65$ at $z = 1$.}
    \label{fig:spatial_energetic}
\end{figure}
%%%%%%
%%%%%%
\subsubsection{Spatial structure of energy exchange terms}
\label{spatial_structure}
%%%%
Here, we examine the spatial structure of the energy exchange terms 
at two Rossby numbers \-- the BCI mode at $Ro=2.5\times10^{-3}$ and the ICL instability mode at $Ro = 1$, both under hydrostatic conditions (figure \ref{fig:spatial_energetic}).
For the BCI mode, HSP is antisymmetric about $y=0$ (figure \ref{fig:spatial_energetic}$(a)$), a consequence of the negative correlation $\langle uv \rangle_x < 0$ (vector plots in figures \ref{fig:eigfuns_cmp}$(b,c)$) and the antisymmetric structure of the horizontal shear $\partial_y U$ (figure \ref{fig:Ro_Ri}$(a)$). As a result, the domain-integrated HSP is close to zero. This antisymmetric structure of HSP is also observed in the non-hydrostatic case (not shown), explaining why the domain-integrated HSP remains negligible for the BCI mode (figure \ref{fig:energetic}$(a)$). In contrast, the VSP remains  positive throughout the domain (figure \ref{fig:energetic}$(b)$), resulting from the negative correlation $\langle uw \rangle_x < 0$ combined with the positive vertical shear $\partial_z U$ (not shown). The value of BFLUX is positive and strongly localized across the cross-front (figure \ref{fig:spatial_energetic}$(c)$).

\begin{figure}
    \centering
    \includegraphics[width=\linewidth]{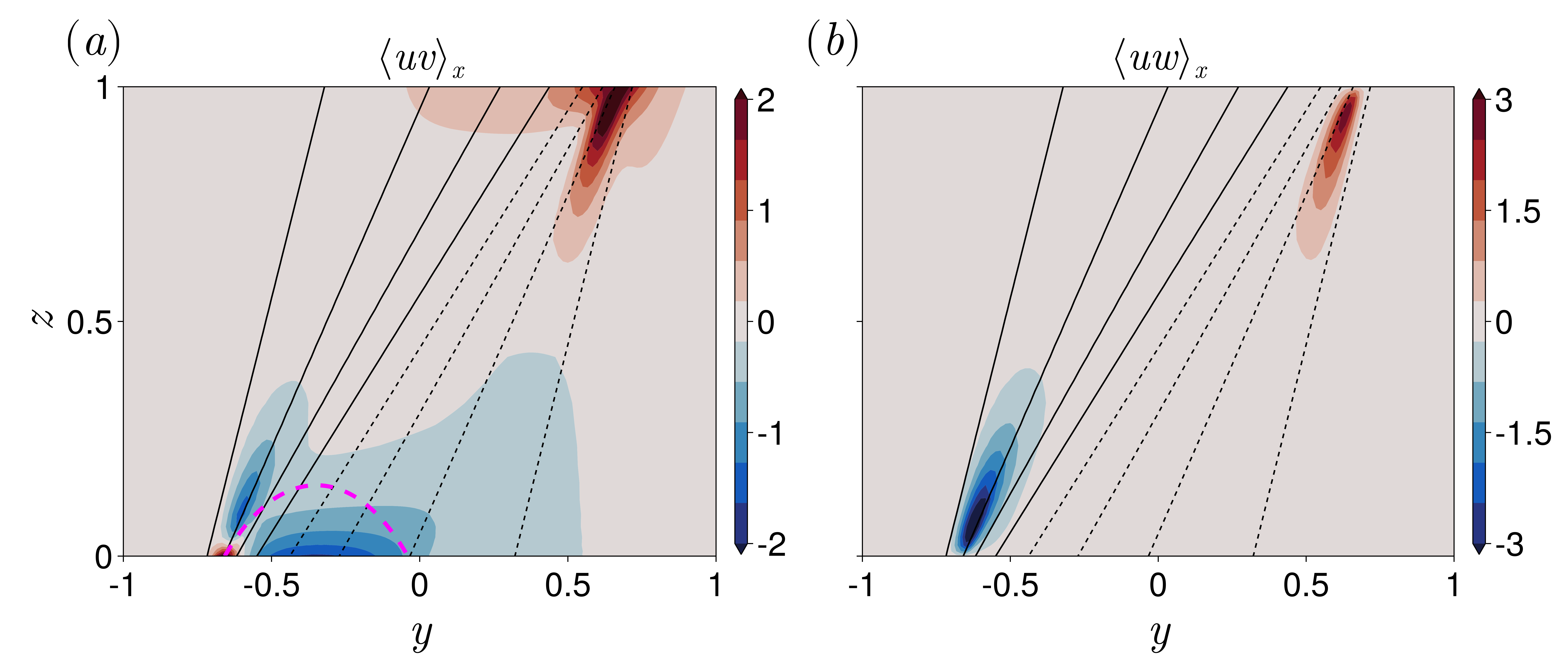}
    \caption{Color contour plots of the $x$-integrated correlation functions for the case $Ro = 1$ under hydrostatic conditions \-- $(a)$ $\langle uv \rangle_x$ and $(b)$ $\langle uw \rangle_x$. The black contours in both panels represent the adjusted isopycnal field $B$, plotted with a contour interval of 0.11; solid (dashed) lines indicate positive (negative) values. In panel (a), the dotted magenta line marks the location of the ICL as defined in (\ref{def_icl}).} 
    \label{fig:flux}
\end{figure}

To further elucidate the energetics of the ICL instability mode and in particular the observed decrease in  $\langle \text{BFLUX} \rangle_{yz}$ with increasing $Ro$ (filled green circle in figures \ref{fig:energetic}$(a,b)$), we examine the spatial structure of the energy exchange terms associated with the positive frequency branch (figures \ref{fig:spatial_energetic}$(d-f)$; the negative frequency branch exhibits a mirrored pattern). 
The positive HSP signal at the upper frontal region, where the vorticity is cyclonic (figure \ref{fig:spatial_energetic}$(d)$), is associated with the Rossby wave (figure \ref{fig:eigfuns_cmp}$(k)$). This arises as a result of the positive correlation between $u$ and $v$ in the upper part of the domain ($\langle uv \rangle_x > 0$; vector plots in figure \ref{fig:eigfuns_cmp}$(k)$ and figure \ref{fig:flux}$(a)$).
%%%%
At the lower frontal region HSP is positive near $z=0$ and then changes sign from being positive to negative (figure \ref{fig:spatial_energetic}$(a)$). This is because  $\langle uv \rangle_x$ also changes sign from positive to negative with increasing $z$ near the ICL region (figure \ref{fig:flux}$(a)$). This sign reversal of $\langle uv \rangle_x$ is an intrinsic property of an IGW that crosses the ICL \citep[shown by dotted magenta line in figure \ref{fig:flux}$(a)$;][]{maslowe1986critical}, and occurs because there is a $\pi/2$ phase shift in the IGW polarization relations between $u$ and $v$. 
%%%
The magnitude of VSP for the ICL instability mode peaks at the top and bottom frontal regions (figure \ref{fig:spatial_energetic}$(b)$), where the vertical shear is strong and positive (not shown). The negative (positive) VSP values near the top (bottom) frontal regions are associated with   $\langle uw \rangle_x>0$ ($\langle uw \rangle_x<0$; figure \ref{fig:flux}$(b)$). 
%%%%
BFLUX exhibits a sign change across the domain (figure \ref{fig:spatial_energetic}$(c)$). Positive BFLUX values, associated with the Rossby wave in the upper domain, indicate the conversion of potential energy into kinetic energy. In contrast, negative BFLUX values are found in the lower frontal region and are related to the Doppler-shifted IGW. 

\begin{figure}
    \centering
    \includegraphics[width=\linewidth]{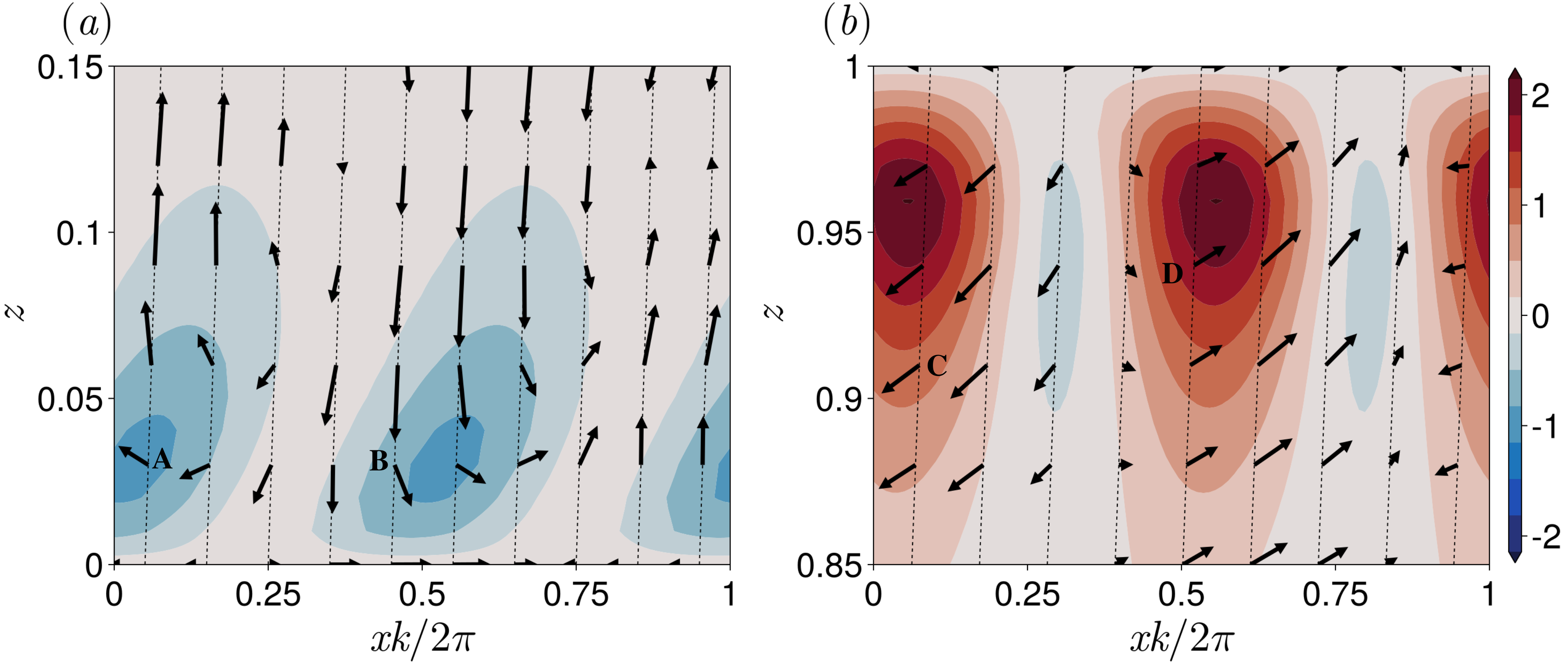}
    \caption{Vector plots illustrating parcel displacements in the $y$- and $z$-directions, represented by $\xi = Dv/Dt$ and $\eta = \epsilon\, Dw/Dt$, respectively, are shown for the case of $Ro = 1$ under hydrostatic conditions. Color contours display $wb$ along the $x$-axis. Panels $(a)$ and $(b)$ correspond to cross-sections at $y = y_\text{ICL}$, $z = 0$, and $y = y_\text{front}$, $z = 0$, respectively (cf. figure \ref{fig:energetic}($f$)). Dashed lines in both panels indicate the angle of the adjusted isopycnals. The $x$-axis is normalized by the perturbation wavelength, $2\pi/k$. Only the lower and upper portions of the domain where the buoyancy flux is concentrated are shown in panels $(a)$ and $(b)$, respectively.
    }
    \label{fig:parcel}
\end{figure}

The spatial structure of BFLUX can be analysed using the parcel method, a conceptual approach used to understand the mechanism of BCI \citep{thorpe1989parcel}. The flow is considered baroclinically unstable if a parcel is displaced adiabatically within the wedge of instability \-— defined as the region between the horizontal layer and the isopycnal slope. For instance, parcels labeled A and B experience instantaneous displacements that exceed the isopycnal slope, resulting in a negative BFLUX (see figure \ref{fig:parcel}$(a)$). Conversely, parcels C and D undergo displacements that remain within the wedge of instability, leading to a positive BFLUX (see figure \ref{fig:parcel}$(b)$). The magnitude of the negative BFLUX increases with increasing $Ro$, thereby explaining the observed decreases in $\langle \text{BFLUX} \rangle_{yz}$.

%%%%
% \begin{figure}
%     \centering
%     \includegraphics[width=\linewidth]{wb_ep0.1_beta2.0_posFreq.png}
%     \caption{Color contour plot of the perturbation vertical vorticity in the $x-z$ plane along $(a)$ $y=y_\text{ICL}$
%     %which is located at $y=-0.66$ 
%     (shown by the thick magenta line in figure \ref{fig:energetic_contour}$(a)$) and $(b)$ $y=y_\text{front}$, which is at location of the maximum magnitude of $\partial_y B$ at $z=1$ (black line in figure \ref{fig:mean_flow}$(d)$). The green line in all the panels shows the perturbation $b$ with a contour internal value of $0.4$, with the solid (dashed) line denoting a positive (negative) value. The black arrows in panels $(a,b)$ show the velocity vector $(u, \epsilon w)$, normalized by their corresponding maximum velocity amplitudes. The $x-$axis in panels $(a,b)$ is normalized by the wavelength $2\pi/k$.}
%     \label{fig:wb_beta2}
% \end{figure}
%%%%
%%%%

% \begin{figure}
%     \centering
%     \includegraphics[width=\linewidth]{scale_energetic_Rorms_mod.pdf}
%     \caption{$(a)$ \red{Non-dimensional wavelength of the most unstable mode $L^\text{ms}=2\pi/k^\text{ms}$ (non-dimensionalized by $R$; see (\ref{len_scale}$(a)$)) I would literally right $L_{ms}/R$ in the y axis of this plot. It would be much clearer}, as function frontal Rossby number $Ro_\text{front}$ for the hydrostatic case. $(b)$ Various energy exchange terms in (\ref{pertb_ke}), averaged over the upper frontal region, which is defined as the $90$th percentile of horizontal buoyancy gradient magnitudes, $|\partial_y B|$, as a function of $Ro_\text{front}$ for the hydrostatic case. For the ICL modes ($\beta \in [0.1,2]$), the energetics in panel $(b)$ are computed from a linear superposition of the positive and negative frequency modes. The open (filled) diamond in panel $(a)$ displays the BCI (ICL) mode. The top $x$-axis of panel $(b)$ shows inverse of the frontal thickness $1/\delta_\text{front}$ (e.g., figures \ref{mean_flow}$(c,d)$). Each data point in panels $(a,b)$ corresponds to a distinct value of $\beta$, which increases from left to right. The $x$-axes in both panels are shown on a logarithmic scale.}
%     \label{fig:energetic_Ro}
% \end{figure}
%%%%%%
%%%%
\begin{figure}
    \centering
    \includegraphics[width=\linewidth]{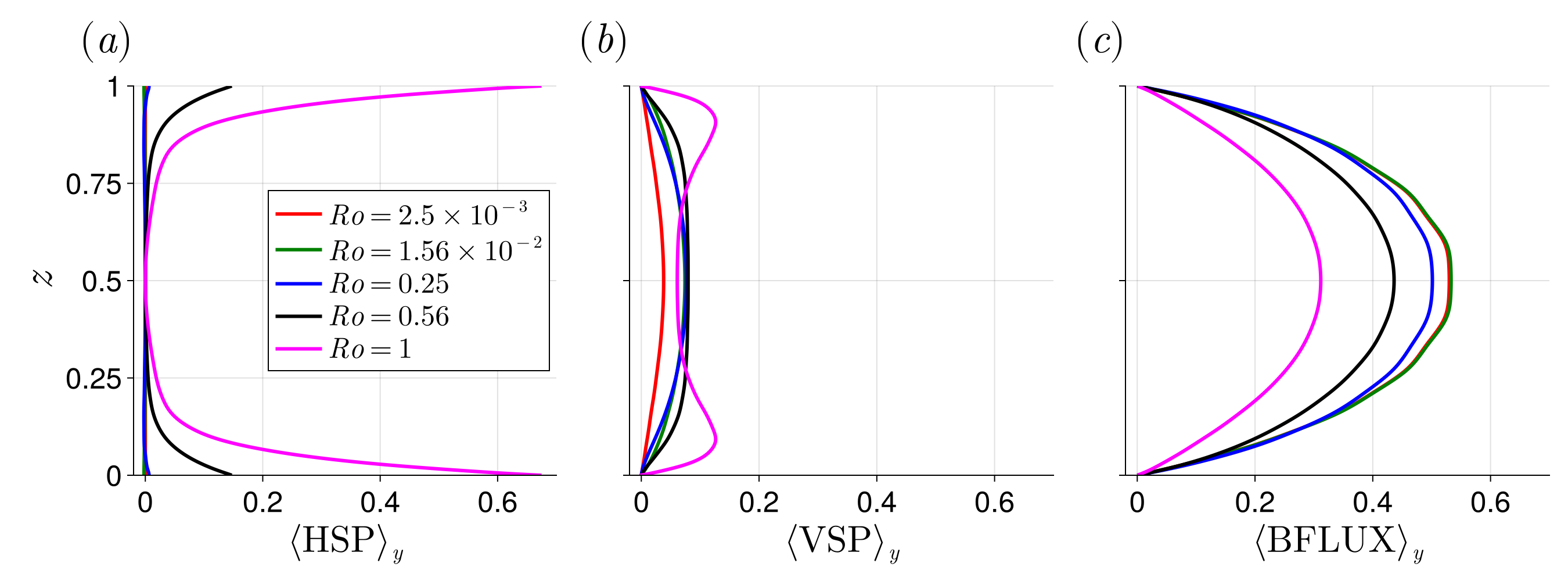}
    \caption{The vertical profiles of the energy exchange terms in (\ref{pertb_ke}) for different values of $Ro$ in the hydrostatic case. 
    %For the ICL modes ($\beta \in [0.5,2]$) the energetics are computed from a linear superposition of the positive and negative frequency modes.
    }
    \label{fig:fullsol}
\end{figure}
\begin{figure}
    \centering
    \includegraphics[width=0.8\linewidth]{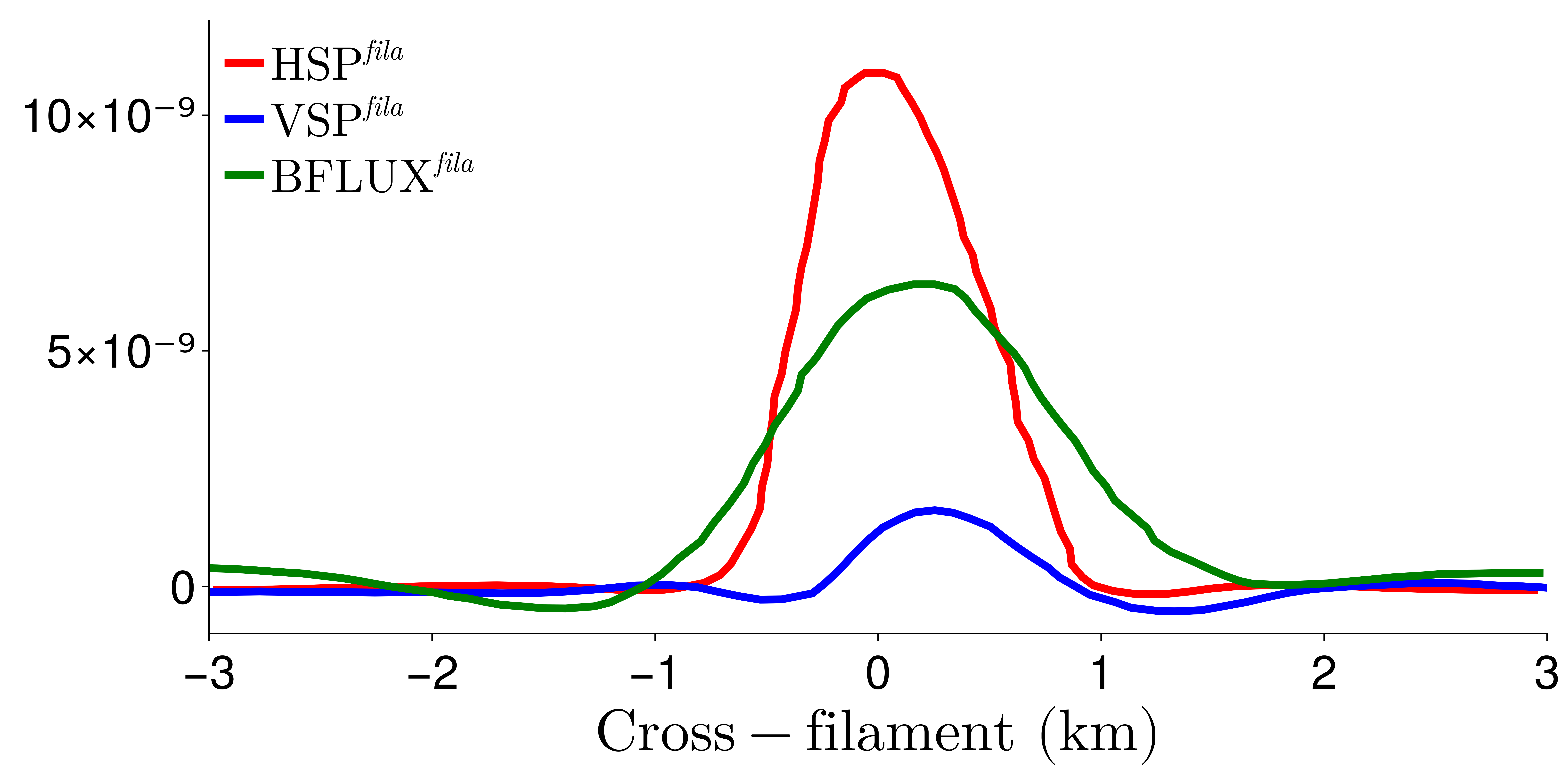}
    \caption{Perturbation KE analysis of a frontal instability in a high-resolution ($150$m horizontal grid spacing) realistic simulation in the Mississippi River plume system, adapted from \cite{wang2021structure}. Mean flow quantities are defined as along-front averages, while perturbation quantities represents deviations from this mean. Based on these decomposition, the exchange terms are \-- $\text{HSP}^\text{fila}=-\overline{{u^\prime}^2} {\partial \overline{u}}/{\partial x} - \overline{{v^\prime}^2} {\partial \overline{v}}/{\partial y} - \overline{u^\prime v^\prime} ({\partial \overline{u}}/{\partial y} + {\partial \overline{v}}/{\partial x})$, $\text{VSP}^\text{fila}=-\overline{u^\prime w^\prime} {\partial \overline{u}}/{\partial z} - \overline{v^\prime w^\prime} {\partial \overline{v}}/{\partial z}$, and $\text{BFLUX}^\text{fila} = \overline{w^\prime b^\prime}$, which are averaged over the top $10$m. The physical interpretation of these terms remains the same as discussed in (\ref{pertb_ke}). The overbar denotes an along-front mean, and the prime denotes a perturbation quantity.}
    \label{fig:filament_instab}
\end{figure}

\section{Discussion}
\label{discussion}
%%%
\subsection{Contrasting basic states: S71 and Ou84 in low-$Ro$ limit}
\label{Ou_vs_Stone}
The remarkable agreement of the growth rates and buoyancy flux magnitudes between S71 and Ou84 for $Ro=2.5\times 10^{-3}$ (figures \ref{fig:growth_rate}(e) and \ref{fig:energetic}(b)) may give the false impression that the Ou basic state in the low-$Ro$ broad-front limit (\ref{ou1984_sol}$a–c$) approaches that of \citet[][(\ref{stone_meanflow}$a,b$)]{stone1971baroclinic}, whereas they are, in fact, fundamentally different. For small ${Ro}$, the isopycnals in Ou84 adjust very slightly (figure \ref{fig:mean_flow}$(c)$) and from (\ref{ou1984_sol}$(a)$) we get 
\begin{align}
    \eta \approx  y \blue{-} Ro^{1/2} \Big(z-\frac{1}{2} \Big).
\end{align}
With ${Ro}^{1/2} \eta \ll 1$, $\tanh(\beta \eta)\approx \beta \eta$ and (\ref{buoy_front}), (\ref{ou1984_sol}$c$) reduce to
%In this section, we demonstrate that the basic state adopted by Stone (\ref{stone_meanflow}$a,b$) differs fundamentally from that obtained by Ou in the low-$Ro$ broad-front limit (\ref{ou1984_sol}$a–c$). For small ${Ro}$, the isopycnals adjust only weakly (figure~\ref{fig:mean_flow}c), yielding $\eta \approx y$.  
%In the subsequent limit $\beta y \ll 1$ leads to $\tanh(\beta y)\approx \beta y$, and Ou’s solution ((\ref{buoy_front}) and (\ref{ou1984_sol}$c$)) reduces to  
\begin{subequations}
\label{ou_lowRo}
\begin{align}
    B(y,z) &\approx -{Ro}^{1/2} \, y + Ro \, \Big(z-\frac{1}{2} \Big), \\
    U(y,z) &\approx {Ro}^{1/2} \left(z - \frac{1}{2} \right), 
\end{align}    
\end{subequations}
which satisfies the thermal-wind relation 
($\partial_z U = -\partial_y B$).  Interestingly, the local Richardson number of the basic state $Ri_\ell=\partial_z B/(\partial_z U)^2\sim1$ because $\partial_z B$ and $(\partial_z U)^2$ both scale with $Ro$ (consistent with figure \ref{fig:Ro_Ri}$(b)$). Thermal wind balance is also satisfied for the \citet{stone1971baroclinic} basic state (\ref{stone_meanflow}$a,b$), but $Ri\sim 1$ only occurs when $\partial_z B\sim\partial_z U\sim1$. 

\subsection{Length scales and energetics}
The {dimensional} wavelength of the most unstable mode, $L^\text{ms}$, increases with $Ro$ under both hydrostatic and non-hydrostatic cases (figures \ref{fig:growth_rate}$(a,d)$, $L^\text{ms}/R \approx 4.2Ro^{1/3},\, 3.9{Ro}^{1/2}$ for non-hydrostatic and hydrostatic cases, respectively), emphasizing the influence of horizontal shear on the instability characteristics. % ($L^\text{ms}/R \approx 4.2Ro^{1/3}$ and $\approx 4.2Ro^{1/3}$ for non-hydrostatic and hydrostatic cases, respectively). %while in the hydrostatic case it scales roughly as $L^\text{ms}/R \approx 3.9{Ro}^{1/2}$ while in the hydrostatic case it scales roughly as $L^\text{ms}/R \approx 3.9{Ro}^{1/2}$ ). Specifically, for the non-hydrostatic case, the relationship follows approximately $L^\text{ms}/R \approx 4.2Ro^{1/3}$, while in the hydrostatic case it scales roughly as $L^\text{ms}/R \approx 3.9{Ro}^{1/2}$ (figures \ref{fig:growth_rate}$(a,d)$). 
For $Ro =1$, $L^\text{ms} \approx 4R$. Using representative oceanic ML parameters in the mid-latitude during winter (e.g., $H=100$ m, $f=10^{-4}$ s$^{-1}$ and $N=10f$ s$^{-1}$) yields a deformation radius $R=1$ km and $L^\text{ms} \approx 4$ km, consistent with the length scale of submesoscale frontal instabilities \citep{wang2021structure}. 

% Specifically, in the non-hydrostatic case, $L^\text{ms}/R$ increases with approximately $4.2Ro^{1/3}$, while for the hydrostatic case, it varies approximately as $3.9\sqrt{Ro}$ (figures \ref{fig:growth_rate}$(a,d)$). This implies that at $Ro \sim \mathcal{O}(1)$, the length scale of the most unstable mode is to be approximately $4R$. Using representative ML depth $H=100$ km and $N=10f$ with $f=10^{-4}$ s$^{-1}$, yields $R=1$ km; thus the length scale of the most unstable is approximately order of $4$ km.

%However, as the reason described in \S \ref{Ou_vs_Stone}, the Ou basic state at low-$Ro$ broad-front case differs from the Stone case; thus we do not expect the scale of the most unstable MLI to be the same. 

% The wavelength of the most unstable mode, $L^\text{ms}$, exhibits only modest variation beyond $Ro_\text{front} > 0.3$ (figure \ref{fig:energetic_Ro}(a)). For instance, $L^\text{ms}$ increases from $2.94$ to $3.69$ as $Ro_\text{front}$ increases from $0.28$ to $0.88$ (figure \ref{fig:energetic_Ro}$(a)$).
% The scale of instability remains below the suggested MLI range for all $\beta$ values. 

As expected for the BCI regime, the dominant source of perturbation energy is the buoyancy flux (open green symbol in figure \ref{fig:energetic}). As $Ro$ increases, the vertical and horizontal shear production grow and their sum becomes comparable to the buoyancy flux when $Ro=1$ (orange and green symbols in figure \ref{fig:energetic}). For larger $Ro$, we anticipate that the shear production terms will continue to grow while the buoyancy flux will continue to weaken, making the shear production terms the dominant source of perturbation energy for $Ro \gg 1$. We note in passing that in the oceanic ML, the vertical shear is likely weaker than in our stability analysis due to the continuous vertical mixing of momentum, and so the stability analysis may overestimate the vertical shear production.

\citet{dong2020scale} estimated that the typical length scale of the most unstable MLI mode is about $6$ km at mid-latitudes, implying that a horizontal grid spacing of order $1$ km is needed to resolve it in oceanic regional models. At this resolution, the local Rossby number is typically $\mathcal{O}(1)$ \citep{capet2008mesoscale1,barkan2017submesoscale}. 
Therefore, the shear production terms (mainly HSP) are expected to significantly affect MLI characteristics at this scale.
% ---- a dynamical process that is entirely overlooked.   

%HSP becomes dynamically significant precisely at the resolutions targeted by the submesoscale-permitting oceanic models. 

% \blue{Previously you had a discussion of TTW balance and Abigail's paper. It wasn't clear to me how it fits here, but I am not against it. }
% Recent efforts to incorporate MLI-induced restratification parametrization into coarse-resolution climate models \-- most notably by \citet{bodner2023modifying} have proposed  that link arrested frontal width to surface-forced turbulent thermal wind (TTW) balance. These frameworks offer significant improvements over earlier MLE closures by introducing mechanistic scaling laws derived from energetically consistent balances. However, the present results indicate that existing parameterizations may under-represent the role of horizontal shear, particularly in regimes where $Ro \gg 1$ and therefore may miss critical energy pathways in the dynamics. This omission may lead to an underestimation of the mixing and restratification potential of strongly ageostrophic fronts. As such, advancing parameterizations to include both TTW dynamics and shear-instability energetics is essential for improving the fidelity of submesoscale physics in global ocean circulation models.

Standard efforts to parametrize the effects of MLI in coarse-resolution climate models \citep[e.g.,][]{fox2008parameterization, bodner2023modifying} focus on the restratification effects associated with the buoyancy flux and do not include the horizontal or vertical shear production. Our work suggests that when $Ro \ge 1$, existing parametrizations may miss critical energy pathways that are essential for adequately representing submesoscale physics in global ocean circulation models.

%Recent efforts to parameterize ML restratification in coarse-resolution climate models \-- most notably the scheme of \citet{bodner2023modifying} that links the arrested frontal width to surface-forced turbulent thermal-wind (TTW) balance. These frameworks improve on earlier MLE closures by introducing mechanistic, energetically consistent scaling laws. The present study, however, indicates that existing schemes under-represent horizontal-shear production when $Ro \gg 1$ and may therefore miss critical energy pathways. This omission may lead to an underestimation of the mixing and restratification of strongly ageostrophic fronts. Advancing closures to couple TTW dynamics with shear-instability energetics is thus essential for faithfully representing submesoscale physics in global ocean circulation models.

\subsection{Frontal arrest}

In realistic oceanic settings, frontogenesis sharpens horizontal buoyancy gradients and horizontal shear at a super-exponential rate \citep{barkan2019role}, a process tending towards a finite-time singularity in the inviscid limit. Ultimately, an instability mechanism may arrest this sharpening before the singularity is reached \citep[e.g.,][]{sullivan2024oceanic}. At the arrested stage, fronts typically exhibit  $Ro\gg1$, and the vorticity field becomes strongly asymmetric, characterized by intense, spatially localized cyclonic vorticity with broader and weaker anticyclonic counterparts.

Although the basic state in our solution does not fully attain such high Rossby numbers, it captures these salient features already at $Ro=1$ (figure \ref{fig:Ro_Ri}$(c)$). At this Rossby number, the horizontal shear production is large and strongly confined near the upper and lower frontal flanks, while the vertical shear production is only marginally stronger than for lower Rossby number results (figures \ref{fig:fullsol}$(a,b)$). In contrast, the buoyancy flux weakens with increasing $Ro$ and remains symmetric with respect to the domain mid-depth. At these frontal regions, the perturbation KE budget is dominated by the horizontal shear production and buoyancy flux, with the vertical shear production contributing marginally (typical values averaged over regions exceeding the $80$th percentile of horizontal buoyancy gradients are $\text{HSP}=0.15$, $\text{VSP}=0.05$, and $\text{BFLUX}=0.13$ for the horizontal shear production, vertical shear production, and buoyancy flux, respectively).
This energetic signature closely matches previously reported results from a submesoscale-resolving realistic-ocean simulation by \citet{wang2021structure}, where the frontal arrest was driven predominantly by the horizontal shear production and buoyancy flux, with the vertical shear production remaining slightly positive (figure~\ref{fig:filament_instab}). A similar pattern emerges in the large-eddy simulation study of arrested filament evolution by \citet{sullivan2018frontogenesis}, who linked frontal arrest to horizontal shear instability.

\section{Summary}
\label{summary}
In this study, we perform a bi-global linear stability analysis of geostrophically adjusted $2$D fronts with zero PV, archetypical of weakly stratified oceanic mixed layers. The frontal configuration follows the analytical solution developed by \citet{ou1984geostrophic}, allowing for a systematic examination of frontal stability characteristics across a broad range of Rossby numbers. 

For low Rossby numbers, the most unstable mode resembles that of classical baroclinic instability. Namely, its growth rate agrees closely with the prediction of \citet{stone1971baroclinic}, and the source of perturbation growth comes solely from the buoyancy flux. Despite these similarities, the low Rossby number  \citet{ou1984geostrophic} basic state describes a different dynamical regime than that investigated by \citet{stone1971baroclinic}.

As the Rossby number increases, the dominant instability transitions to an inertia-critical layer (ICL) instability mode, with conjugate pairs of positive and negative frequencies exhibiting equal growth rates. This instability results from a resonance between a Rossby wave supported by the isopycnal potential vorticity gradient and an inertia-gravity wave. 

At the order-one Rossby number regime, the buoyancy flux magnitude is comparable to the sum of vertical and horizontal shear production, with the latter dominating near the region where the adjusted front is the strongest. In this regime, the instability agrees with the frontal arrest mechanism discussed by \citet{sullivan2018frontogenesis}, whereby frontal intensification is halted by horizontal shear instability.

% \blue{Although the QG model yields higher growth rates than the primitive-equation model, horizontal shear reverses the energy pathway: in QG it siphons energy from the perturbations to the mean flow, whereas in the primitive equations it feeds energy from the mean flow back into the disturbances.}

Mixed-layer instability (MLI) is traditionally viewed as a purely baroclinic process that restratifies the ML by releasing the available potential energy stored in horizontal buoyancy gradients \citep{boccaletti2007mixed}. The present stability analysis suggests that MLI can exhibit features of ICL instability, with a comparable growth rate to the mixed layer baroclinic instability but exhibiting both baroclinic and barotropic characteristics. This can have important implications for submesoscale parametrizations in ocean models.

Existing MLI parameterization schemes that are based on 
$Ro\ll 1, \, Ri\sim \mathcal{O}(1)$
baroclinic instability theory may fail to capture key energetic processes in realistic, ageostrophic fronts. In particular, horizontal shear production, which is significant in the  R$o\sim \mathcal{O}(1)$ regime, is completely ignored in mixed layer eddy parameterizations \citep[e.g.,][]{fox2008parameterization}. This highlights the need for revised parametrization frameworks that incorporate both baroclinic and barotropic instability mechanisms, particularly under conditions of strong frontal intensification, to more accurately represent upper-ocean restratification and kinetic energy dissipation.

\clearpage
\section*{Acknowledgements}
SK and RB are supported by ISF grant 2054/23 and by the European Union (ERC, 401 ML Transport, 101163887). Views and opinions expressed are, however, those of the author(s) only and do not necessarily reflect those of the European Union or the
European Research Council. Neither the European Union nor the granting authority can be held responsible for them.

\section*{Declaration of Interests}
The authors report no conflicts of interest.

\section*{Data availability statement}
The linear stability code used in this study is available 
at \url{https://github.com/subhk/Frontal_Stability}.

%\FloatBarrier
%\clearpage
\appendix
\section{Matrices of the generalized eigenvalue problem}
\label{matrices_details}
The elements of $\bm{\mathsfit{A}}$ and $\bm{\mathsf{B}}$ of 
(\ref{gen_eigvals}) are given by
\begin{subequations}
\label{eigs_matrix}
\begin{align}
    \bm{\mathsfit{A}} &= \begin{pmatrix}
        a_{11} & a_{12} & a_{13} \\
        a_{21} & a_{22} & 0 \\
        a_{31} & (\ii k  \partial_y B \mathsfit{I}) \mathsfit{H} & 
        \ii k U \mathsfit{I} - E \mathsfit{D}^2
    \end{pmatrix},
    % \\ \nonumber \\ 
    % \bm{\mathsf{X}} &= \begin{pmatrix}
    %     \tilde{w} \\ \tilde{\zeta} \\ \tilde{b}
    % \end{pmatrix},
\,\,\,\,\ %\\\\
    \bm{\mathsfit{B}} = \begin{pmatrix}
        -\mathsfit{D}^2 & 0 & 0 \\
        0 & -\mathsfit{I} & 0 \\
        0 & 0 & -\mathsfit{I}
    \end{pmatrix},
\tag{B1$b,c$} 
\end{align}    
\end{subequations}
where
\begin{subequations}
\begin{align}
    a_{11} &= (\ii k U \mathsfit{I}) \mathsfit{D}^2 - E \mathsfit{D}^4 + \ii k \bigg(\partial^2_y {U} - \frac{1}{\epsilon^2} \partial^2_z {U} \bigg) \mathsfit{I}
    + (2 \ii k \partial_y U \mathsfit{I}) (\mathsfit{D}_y \otimes \mathsfit{I}_z)
    \nonumber \\
    &+ \bigg(\frac{2}{\epsilon^2} \ii k \partial_{y} U \mathsfit{I} \bigg) \mathsfit{H} (\mathsfit{D}_y \otimes \mathsfit{D}^2_z) 
    + {\bigg(\frac{2}{\epsilon^2} \ii k \partial_{yz} U \mathsfit{I} \bigg) \mathsfit{H} (\mathsfit{D}_y \otimes \mathsfit{D}_z)}, 
\\\
    a_{12} &= \frac{1}{\epsilon^2} \mathsfit{I}_y \otimes \mathsfit{D}_z 
    + \bigg(\frac{2}{\epsilon^2} k^2 \partial_y U \mathsfit{I} \bigg) \mathsfit{H} (\mathsfit{I}_y \otimes \mathsfit{D}_z )
    + {\bigg(\frac{2}{\epsilon^2} k^2 \partial_{yz} U \mathsfit{I} \bigg)
    \mathsfit{H}} ,
\\\
    a_{13} &= -\frac{1}{\epsilon^2} 
    (\mathsfit{D}_y^2 \otimes \mathsfit{I}_z - k^2 \mathsfit{I})
\\\
    a_{21} &= -\partial_{yz} U \mathsfit{I} - \partial_z U \mathsfit{I} (\mathsfit{D}_y \otimes \mathsfit{I}_z)
    + \partial_y U \mathsfit{I} (\mathsfit{I}_y \otimes \mathsfit{D}_z) -  (\mathsfit{I}_y \otimes \mathsfit{D}_z) 
    \nonumber \\
    &+ (\partial^2_y U \mathsfit{I}) \mathsfit{H} (\mathsfit{D}_y \otimes \mathsfit{D}_z),
\\\    
    a_{22} &= ik U \mathsfit{I} - E \mathsfit{D}^2
    -(\ii k \partial^2_y U \mathsfit{I}) \mathsfit{H},
\\\
    a_{31} &= \partial_z B \mathsfit{I} - (\partial_y B \mathsfit{I}) \mathsfit{H} (\mathsfit{D}_y \otimes \mathsfit{D}_z),
\end{align}
\end{subequations}
%%%%
where $\otimes$ is the Kronecker product. $\mathsfit{I}_y$ and $\mathsfit{I}_z$ are identity matrices of size $(N_y \times N_y)$ and $(N_z \times N_z)$ respectively, and $\mathsfit{I}=\mathsfit{I}_y \otimes \mathsfit{I}_z$.
%has size of $(N_y \times N_z)^2$.
The differential operator matrices are given by
\begin{subequations}
\begin{align}
    \mathsfit{D}^2 &= \frac{1}{\epsilon^2} \mathsfit{I}_y \otimes \mathsfit{D}_z^2
    + \mathsfit{D}_y^2 \otimes \mathsfit{I}_z - k^2 \mathsfit{I}, 
\\\
    \mathsfit{D}^4 &= \frac{1}{\epsilon^4} \mathsfit{I}_y \otimes \mathsfit{D}_z^4 + \mathsfit{D}_y^4 \otimes \mathsfit{I}_z 
     + k^4 \mathsfit{I} - 2k^2 \mathsfit{D}_y^2 \otimes \mathsfit{I}_z -\frac{2k^2}{\epsilon^2}\mathsfit{I}_y \otimes \mathsfit{D}_z^2 + \frac{2}{\epsilon^2} \mathsfit{D}_y^2 \otimes \mathsfit{D}_z^2, 
\\\
    \mathsfit{H} &= (\mathsfit{D}_y^2 \otimes \mathsfit{I}_z - k^2 \mathsfit{I})^{-1}, 
\end{align}    
\end{subequations}
where $\mathsfit{H}$ describes the inverse of the horizontal Laplacian operator. 
%$(\mathcal{D}_h^2)^{-1}$.
%%%%
% A standard approach is followed in the construction of differentiation matrices \citep{trefethen2000spectral}. The transformed Gauss–Lobatto points for $z \in [0, 1]$ are given by
% \begin{align}
%     z_j = \frac{1}{2} \cos{(j\pi/N_z)} + \frac{1}{2},
%     \,\,\,\,
%     j = 0, \cdots, N_z.
% \end{align}
% and the first-order Chebyshev differentiation matrix is given by
% \begin{equation}
%   (\mathsfit{D}_z)_{ij} = \begin{cases}
%      \dfrac{2N_z^2+1}{3},  \,\,\,\,\, i=j=0, 
% \\
%     \dfrac{c_i}{c_j} \dfrac{(-1)^{i+j}}{z_i-z_j},
%      \,\,\,\,\, i \neq j,
%      \,\,\,\,\, c_i
%      \begin{cases} 
%         2, \,\,\,\,\, i=0,N_z, \\
%         1, \,\,\,\,\, \text{otherwise},
%      \end{cases}
% \\
%     \dfrac{-\cos{(j\pi/N_z)}}{1-\cos^2{(j\pi/N_z)}},
%     \,\,\,\,\, 0 < i = j < N_z,
% \\
%     -\dfrac{2N_z^2+1}{3}, \,\,\,\,\, i=j=N_z.
%   \end{cases}
% \end{equation}
% For $y \in [0,L_y]$, the first-order Fourier differentiation matrix for even $N_y$ is,
% \begin{equation}
%     (\mathsfit{D}_y)_{ij} = \begin{cases}
%         0, \,\,\,\,\, i=j, 
% \\
%         \dfrac{\pi}{L_y} (-1)^{i-j} \cot{\left(\dfrac{(i-j)h}{2} \right)}
%         \,\,\,\,\, i \neq j,
%     \end{cases}
% \end{equation}
% and for odd $N_y$,
% \begin{equation}
%     (\mathsfit{D}_y)_{ij} = \begin{cases}
%         0, \,\,\,\,\, i=j, 
% \\
%         \dfrac{\pi}{L_y} (-1)^{i-j} \csc{\left(\dfrac{(i-j)h}{2} \right)}
%         \,\,\,\,\, i \neq j,
%     \end{cases}
% \end{equation}
% where $h=2\pi/N_y$.

\section{Benchmark of the stability code}
\label{benchmark}
The linear stability code used in this study is benchmarked against the asymptotic solution for BCI by \citet{stone1971baroclinic}.
The nondimensional form of the basic state is
\begin{subequations}
\label{stone_meanflow}
\begin{align}
    U(y,z) &= z - \frac{1}{2}, \\
    B(y,z) &= Ri z - y.
\end{align}
\end{subequations}
% where $Ri=(N H/U)^2$ is the Richardson number, and $N$ represents the dimensional buoyancy frequency. 
% \cite{stone1966non} derived an analytical approximation for the growth rate as a function of along-front wavenumber $k$ by expanding the eigenvalue problem for small $k$ and $l=0$, where $l$ is the wavenumber in the cross-front $(y)$ direction. The approximate growth rate $\sigma_r$ of the most unstable BCI mode is given by, 
% \begin{align}
% \label{stone_sol}
%     \sigma_r^\text{Stone} \approx  \frac{1}{2\sqrt{3}} 
%     \Big[k - \frac{2k^3}{15}\Big(1 + Ri_s + \frac{5k^2\epsilon^2}{42} \Big) \Big].
% \end{align}
% where $\epsilon = H/\lambda = f/N_s$, where $\lambda$ represents the gravest baroclinic deformation radius \citep{molemaker2005baroclinic}.

\begin{figure}
    \centering
    \includegraphics[width=\textwidth]{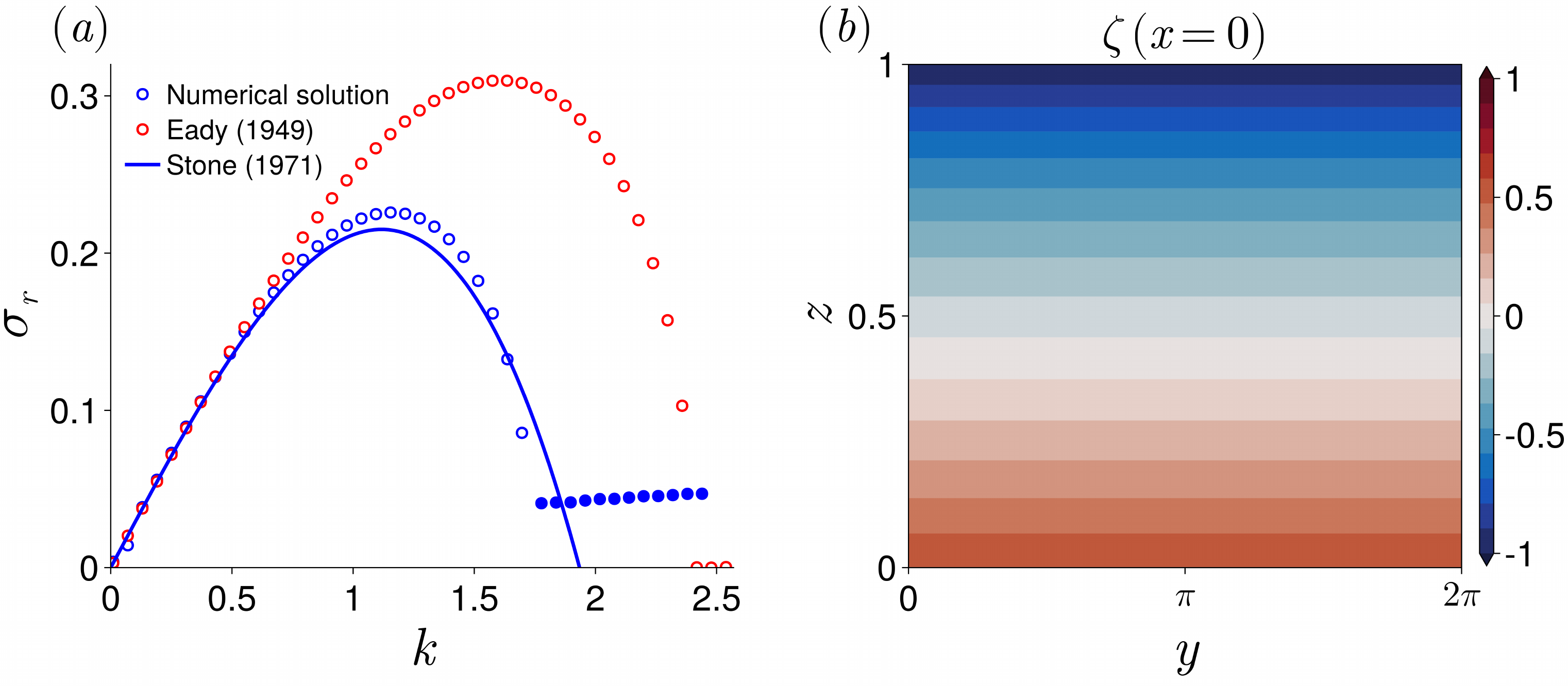}
    \caption{$(a)$ Comparison of numerically obtained growth rate $\sigma_{r}$ (blue circle) with the analytical solution of \citet[][blue line]{stone1971baroclinic} given by (\ref{stone_sol}), for $\epsilon=0.1$ and $Ri=1$. Also shown is the growth rate obtained from the QG approximation \citep[][open red circles; see Appendix \ref{qg_stability}]{eady1949long}. Open blue circle shows the BCI modes, while filled blue circle represent ICL modes. $(b)$ The vertical vorticity $\zeta$ for the most unstable BCI mode ($k=1.16$) from the numerical solution. The domain sizes are $y \in [0, 2\pi]$ and $z \in  [0, 1]$, respectively, and the solutions are obtained using a grid resolution of $N_y=120$ and $N_z=24$. }
    \label{fig:benchmark}
\end{figure}

\cite{stone1966non} derived an analytical approximation for the growth rate as a function of along-front wavenumber $k$ by expanding the eigenvalue problem for small $k$ and $l=0$, where $l$ is the wavenumber in the cross-front $(y)$ direction. The approximate growth rate of the most unstable BCI mode, $\sigma_r^\text{Stone}$, is given by
\begin{align}
\label{stone_sol}
    \sigma_r^\text{Stone} \approx  \frac{1}{2\sqrt{3}} 
    \Big[k - \frac{2k^3}{15}\Big(1 + Ri + \frac{5k^2\epsilon^2}{42} \Big) \Big],
\end{align}
where $\epsilon = H/R$. 
% \blue{how is  R here different than the way you are defining it in the manuscript? }
%%%
Stone’s asymptotic solution (\ref{stone_sol}) with $Ri = 1$ provides a good approximation to the numerical results for $k \lesssim 2$ (open blue circles in figure \ref{fig:benchmark}$(a)$).  In contrast, the solution of \citet{eady1949long} overestimates the growth rate in this regime. The discrepancy arises because, at $Ri=1$, ageostrophic effects become non-negligible and are not captured by the QG framework.
%%%
Unlike Stone’s solution, the numerical results do not exhibit a short-wave cutoff for instability. 
% \blue{Fig. 13a seems to show short wave cutoff to me in the Eady problem. Am I missing something?}. 
At $k \gtrsim 2$, the perturbation modes have relatively smaller growth rates \citep[filled blue circles in figure \ref{fig:benchmark}$(a)$; see also figure 2 of][]{stone1970non}. This branch of instability arises from a resonant interaction between a Rossby wave and a Doppler-shifted internal gravity wave (IGW) \citep{nakamura1988scale}. Nevertheless, the structure of the most unstable BCI mode remains invariant in the cross-front ($y$) direction, consistent with the theoretical prediction of \citep[][figure \ref{fig:benchmark}$(b)$]{stone1966non}.

\section{Grid independent test of the stability results}
\label{grid_test}
% To avoid the influence of the periodic boundary condition on the solution of the stability analysis, the cross-frontal length is set to $3L_f$, which is large enough since the most unstable modes are confined in the cross-front direction (e.g., figure \ref{fig:eigfun_cmp}$(a-c)$). The grid convergence results are obtained for the cases of $\beta=0.1$ and $\beta=2$ with $N_y=32$ grid points in $y$, $N_z=240$ grid points in $z$ (Table \ref{tab:grid_test}), leading to the sizes of $\bm{\mathsfit{A}}$ and $\bm{\mathsfit{B}}$ being $(3\times N_y \times N_z)^2=(23040)^2$. 
%%%%
Grid resolution convergence tests for the most unstable mode were conducted for two extreme values of $Ro$ \-- $Ro=2.5\times10^{-3}$ and $Ro=1$. The growth rate of the most unstable mode converges at $N_y = 240$ and $N_z = 32$ (table \ref{tab:grid_test}). Accordingly, all results presented in this paper are based on $N_y = 240$ and $N_z = 32$.

%%%%%%
%%%%%%
\begin{table}
\centering
\setlength\tabcolsep{20pt}
\caption{Grid convergence tests for the linear stability analysis in the hydrostatic regime were performed for the most unstable mode. Results are shown for two cases \-- $Ro = 2.5\times10^{-3}$ at $k = 20.614$, and $Ro = 1$ at $k = 1.698$ (figure \ref{fig:growthrate_k}). $N_y$ and $N_z$ denote number of grid points in the $y$ and $z$ directions, respectively.
}
\label{tab:grid_test}
\scriptsize
\begin{tabular}{lrrr}
\toprule
%\multirow{2}{*} 
{\small $(N_y \times N_z)$} 
&\multicolumn{1}{c}{\small $\sigma(Ro = 2.5\times10^{-3})$} 
&\multicolumn{1}{c}{\small $\sigma(Ro = 1)$} 
\\\midrule
\small $(24\times120)$ 
&\multicolumn{1}{c}{\small $0.2043$} 
&\multicolumn{1}{c}{\small $0.1872 \pm 0.1053 \ii$} 
\\
\small $(24\times240)$ 
&\multicolumn{1}{c}{\small $0.2043$} 
&\multicolumn{1}{c}{\small $0.1881 \pm 0.1061 \ii$} 
\\
\small $(32\times240)$ 
&\multicolumn{1}{c}{\small $0.2043$} 
&\multicolumn{1}{c}{\small $0.1881 \pm 0.1061 \ii$} 
\\
\bottomrule
\end{tabular}
\end{table}

\section{Formulation of the 2D QG stability analysis}
\label{qg_stability}
In this section, we outline the formulation of the $2$D QG stability problem. The non-dimensional form of the linearized version of the QG PV perturbation equation under the $f$-plane approximation can be expressed as \citep{pedlosky2013geophysical},
\begin{align}
\label{qg_pv_eq}
    \frac{\partial q^\text{qg}}{\partial t} + U \frac{\partial q^\text{qg}}{\partial x} + \frac{\partial \psi}{\partial x}
    \frac{\partial Q^\text{qg}}{\partial y} = E \, \nabla_h^2 q^\text{qg}, \,\,\,\,\,\,\  \text{for} \,\,\, 0 < z <1, 
\end{align}
where $q^\text{qg}$ is the perturbation QG PV, and it is defined as 
\begin{align}
\label{qg_pv_def}
    q^\text{qg} = \nabla_h^2 \psi^\text{qg} + 
    \frac{\partial}{\partial z}
    \left(\frac{1}{N^2} \frac{\partial \psi^\text{qg}}{\partial z}\right),
\end{align}
where $N^2$ describes the stratification profile averaged over the frontal zone, and it is defined as
\begin{align}
\label{qg_N2}
    N^2(z) = \frac{1}{L_f} \int_{-L_f/2}^{L_f/2} \partial_z B \, dy,
\end{align}
where $L_f=Ro^{-1/2}$ is the cross-frontal width.
The variable $\psi^\text{qg}$ describes the QG perturbation streamfunction with $u^\text{qg}=-\partial_y \psi^\text{qg}$ and $v^\text{qg}=\partial_x \psi^\text{qg}$. 
The variable $Q^\text{qg}$ describes the QG PV of the basic state, which is defined as \citep{pedlosky2013geophysical}
\begin{align}
    Q^\text{qg} = -\frac{\partial U}{\partial y} + \frac{\partial}{\partial z}\left(\frac{B}{N^2} \right),
\end{align}
and the cross-front gradient of $Q^\text{qg}$ is defined as
\begin{align}
    \frac{\partial Q^\text{qg}}{\partial y} = -\frac{\partial^2 U}{\partial y^2} - \frac{\partial}{\partial z}\left(\frac{\partial_z U}{N^2} \right).
\end{align}
The linearized perturbation buoyancy equation at the top and the bottom boundary is
\begin{align}
\label{qg_b}
    \frac{\partial b^\text{qg}}{\partial t} + U \frac{\partial b^\text{qg}}{\partial x} + \frac{\partial \psi^\text{qg}}{\partial x}
    \frac{\partial B}{\partial y} = 0,
    \,\,\,\,\,\,\ \text{at} \, z=0 \,\ \text{and} \,\, 1,
\end{align}
where $b^\text{qg}=\partial_z \psi^\text{qg}$.
Next, we seek normal-mode solutions for $\psi^\text{qg}$ and $q^\text{qg}$ in the form of 
\begin{align}
\label{qg_mode}
    [\psi^\text{qg}, q^\text{qg}] = \mathfrak{R}\big([\widetilde{\psi}^\text{qg}, \widetilde{q}^\text{qg}] \big)(y, z) \ee^{\ii kx-\sigma t},    
\end{align}
where $\widetilde{\psi}^\text{qg}$, $\widetilde{q}^\text{qg}$ are the eigenfunctions of $\psi^\text{qg}$ and $q^\text{qg}$, respectively.
Using (\ref{qg_mode}), (\ref{qg_pv_eq}), and (\ref{qg_pv_def}), (\ref{qg_b}) can be expressed in terms of streamfunction $\psi^\text{qg}$, 
\begin{subequations}
\label{qg_evp}
\begin{align}
    [(\sigma + \ii k U) - E] \mathscr{L}\widetilde{\psi}^\text{qg} + \ii k \partial_y Q^\text{qg} \widetilde{\psi}^\text{qg} &= 0, \,\,\,\,\  \text{for} \,\, 0 < z <1, 
\\
    (\sigma + \ii k U_{-})\partial_z \widetilde{\psi}^\text{qg}_{-} + \ii k \partial_y B_{-} \widetilde{\psi}^\text{qg}_{-} &= 0, \,\,\,\,\, \text{at} \,\, z = 0,
\\
    (\sigma + \ii k U_{+})\partial_z \widetilde{\psi}^\text{qg}_{+} + \ii k \partial_y B_{+} \widetilde{\psi}^\text{qg}_{+} &= 0, \,\,\,\,\, \text{at} \,\, z = 1, 
\end{align}
\end{subequations}
where $\mathscr{L}$ is a linear operator defined as
\begin{align}
    \mathscr{L} \equiv \mathcal{D}_h^2 + 
    \frac{\partial}{\partial z} \left(\frac{1}{N^2} 
    \frac{\partial}{\partial z} \right),
\end{align}
and $\mathcal{D}_h^2 = (\partial_y^2 - k^2)$. The subscripts $-,+$ in (\ref{qg_evp}$b,c$) denote the values of the fields at $z=0$ and $z=1$, respectively. The above set of equations can be cast into a generalized eigenvalue problem similar to (\ref{gen_eigvals}). We followed similar numerical techniques as discussed in \S \ref{method} to solve the eigenvalue problem. 

We benchmarked the QG stability solver against the analytical solution of the Eady problem, using the basic state given by (\ref{stone_meanflow}) in Appendix \ref{stone_meanflow}. Following the procedure outlined by \citet{vallis2017atmospheric}, we derived the analytical expression for the growth rate, 
\begin{align}
\label{eady_growthrate}
    \sigma_r^\text{Eady} = \frac{1}{\sqrt{Ri}} \Bigg[\Bigg(\coth{\frac{\mu}{2}} - \frac{\mu}{2} \Bigg) \Bigg(\frac{\mu}{2} - \tanh{\frac{\mu}{2}} \Bigg)\Bigg]^{1/2},
\end{align}
where $\mu = k\sqrt{Ri}$. The wavenumber of the most unstable mode is given by $k_m=1.61/\sqrt{Ri}$. For $Ri=1$, the numerically obtained growth rate at $k_m =1.61$, $\sigma_r \approx 0.31$, which matches with the theoretical prediction (\ref{eady_growthrate}).

\subsection{Structure of mode}
\label{governor}
\begin{figure}
    \centering
    \includegraphics[width=\linewidth]{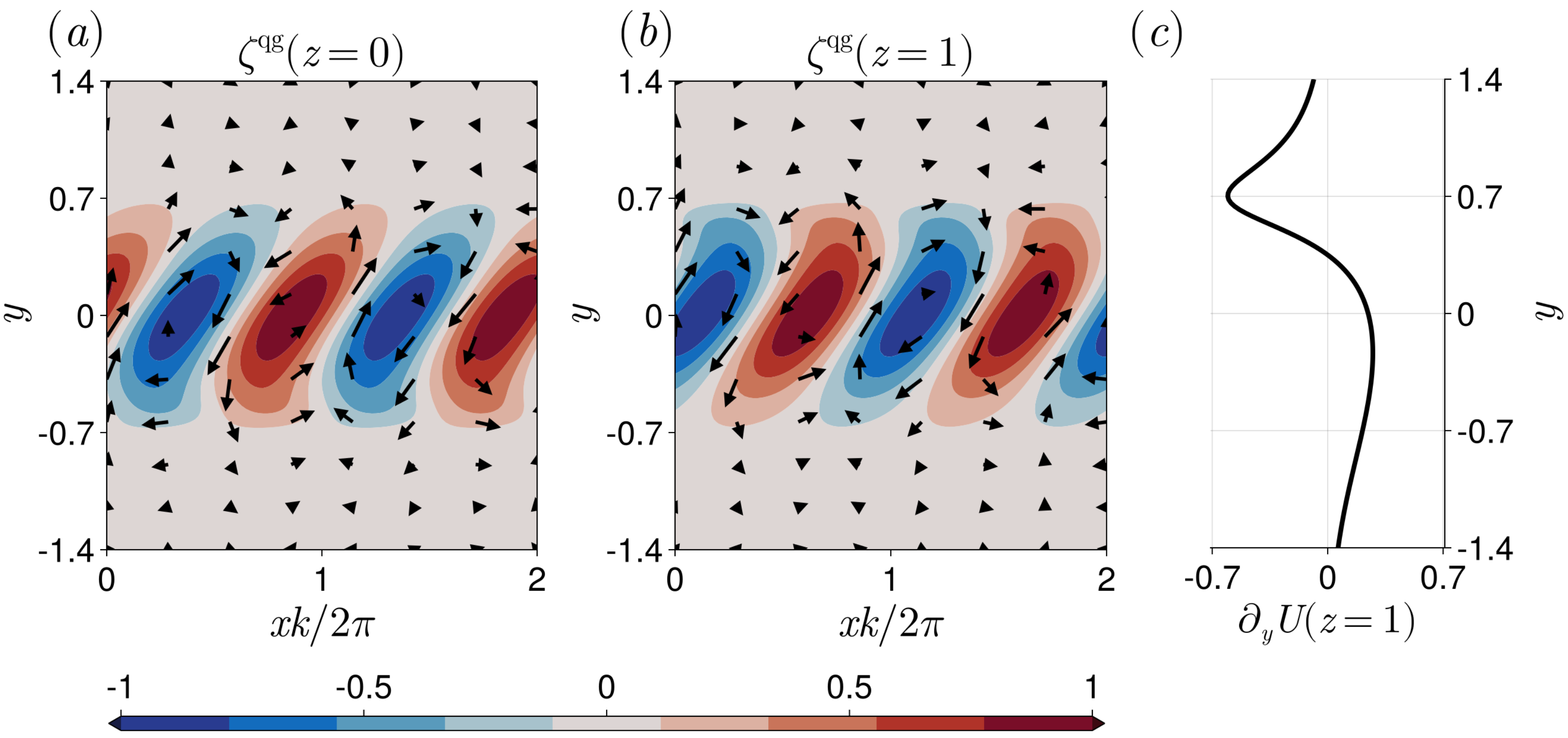}
    \caption{Structure of QG  vertical vorticity $\zeta^\text{qg}(=\nabla_h^2 \psi^\text{qg})$ $(a)$ at $z=0$ and $(b)$ at $z=1$ for the case of $Ro=0.56$ (corresponds to $\beta=1.5$), with superimposed horizontal velocity vector $(u^\text{qg}, v^\text{qg})$ shown by the black arrows. The horizontal circulation in panels $(a,b)$ is clockwise (anticlockwise) around the negative (positive) vorticity region. The velocity vectors in panels $(a)$ and $(b)$ are normalized by their respective maximum velocity amplitudes. 
    The $x-$axis in panels $(b,c)$ are normalized the wavelength $2\pi/k$. Panel $(d)$ shows the horizontal shear $\partial_y U$ at $z=1$.
    }
    \label{fig:eigfun_qg_beta1.5}
\end{figure}

The Rossby wave mode is predominantly localized within the region of anticyclonic vorticity (figures \ref{fig:eigfun_qg_beta1.5}$(a,c)$) and exhibits a rightward tilt (figures \ref{fig:eigfun_qg_beta1.5}$(a,b)$). This spatial structure is indicative of the generation of a counter-gradient horizontal momentum flux, with $\langle u^\text{qg} v^\text{qg} \rangle_x > 0$ , which tends to reinforce the background horizontal shear (figure \ref{fig:eigfun_qg_beta1.5}$(c)$). As a result, $\langle \text{HSP}^\text{qg} \rangle_x$ becomes negative, reflecting a net transfer of energy from the perturbation field to the frontal flow {(red crosses in figure \ref{fig:energetic}b)}.

% The QG BCI mode is predominantly localized in the anticyclonic vorticity region and tilts rightward, reflecting the fact that the mode is producing a counter-gradient
% horizontal momentum flux (i.e., $\langle u^\text{qg} v^\text{qg} \rangle_x > 0$; figures \ref{fig:eigfun_qg_beta1.5}$(a,b)$). 

%This particular feature allows the mode to give
% tilt toward the rightward (figure \ref{fig:eigfun_qg_beta1.5}), demonstrating the effect of horizontal shear on the eigenfunction structure. 

% The phases of QG BCI mode in the horizontal phase tilt more towards rightward as the horizontal shear increases (figure \ref{fig:eigfun_qg_beta1.5}$(b,c)$). 
%%%%
% For the equivalent QG BCI mode, the tilt in the horizontal plane is very less (figure \ref{fig:eigfun_cmp}$(e,f)$). However, this tilt in the phase of the BCI mode is pronounced for larger values of shear (figure \ref{fig:eigfun_qg_beta1.5}$(b,c)$).  
%%%%%
%%%%%

\subsection{QG energetics}
\label{qg_energetic}
In the QG formalism, the evolution of the perturbation KE can be written as \citep{pedlosky2013geophysical}
\begin{align}
\label{qg_ke_eq}
    2 \sigma \langle K^\text{qg} \rangle_x  = \underbrace{\langle-\widetilde{u}^{\text{qg}^\star}
    \widetilde{v}^\text{qg} \frac{\partial {U}}{\partial {y}} \rangle_x}_{\text{HSP}^\text{qg}}
     + \underbrace{\langle \widetilde{w}^{\text{qg}^\star} \widetilde{b}^\text{qg} \rangle_x}_{\text{BFLUX}^\text{qg}}
     + \underbrace{\langle \widetilde{\nabla} \cdot  \left (\widetilde{\bm{u}}^\star \widetilde{p} \right) \rangle_x}_{\text{PWORK}^\text{qg}}
     \nonumber \\
     + \underbrace{\langle E \Big(\widetilde{u}^{\text{qg}^\star} \nabla_h^2 \widetilde{u}^\text{qg} 
     + \widetilde{v}^{\text{qg}^\star} \nabla_h^2 \widetilde{v}^{\text{qg}} 
      \Big) \rangle_x}_{\text{DISP}^\text{qg}},
\end{align}
where $K^\text{qg}$ is the QG perturbation KE, defined as $K^\text{qg}=(\widetilde{u}^\text{qg} \widetilde{u}^{\text{qg}^\star} + \widetilde{v}^\text{qg} \widetilde{v}^{\text{qg}^\star})/2$. 
%%%%
The physical interpretations of horizontal shear production ($\text{HSP}^\text{qg}$), buoyancy flux ($\text{BFLUX}^\text{qg}$), and pressure work ($\text{PWORK}^\text{qg}$) in the QG formalism are identical to those discussed in (\ref{pertb_ke}). The dissipation term ($\text{DISP}^\text{qg}$) is negligible for the unstable perturbations due to the small value of $E$ used in the stability analysis (not shown). The variable $w^\text{qg}$ denotes the vertical velocity in the QG approximation and is obtained by solving the QG $\omega$-equation \citep[see next section]{hoskins1978new}.

\subsection{Calculating vertical velocity}
\label{omgea_equation}
The nondimensional form of the QG $\omega$ equation can be expressed as
\citep{hoskins1978new}
\begin{align}
\label{omgea_eq}
    N^2 \nabla_h^2 w^\text{qg} + \frac{\partial^2 w^\text{qg}}{\partial z^2} = 2 \nabla_h \cdot \bm{\mathcal{Q}},
\end{align}
where $\bm{\mathcal{Q}}$ is defined as
\begin{align}
\label{q_vector}
    \bm{\mathcal{Q}} = -\Bigg(\frac{\partial v^\text{qg}}{\partial x} \frac{\partial B}{\partial y}, \frac{\partial U}{\partial y}
    \frac{\partial b^\text{qg}}{\partial y} + \frac{\partial v^\text{qg}}{\partial y} \frac{\partial B}{\partial y} \Bigg). 
\end{align}
Equation (\ref{omgea_eq}) is solved subject to rigid-lid boundary conditions at $z=0$ and $z=1$, and periodic boundary conditions in the cross-front ($y$) direction.

% An exmaple of 
% %%%%%
% %%%%%
% \begin{figure}
%     \centering
%     \includegraphics[width=\linewidth]{energetic_qg_beta1.5.png}
%     \caption{Color contour plots of $(a)$ $\text{HSP}^\text{qg}$ 
%     and $(v)$ $\text{BFLUX}^\text{qg}$ of the perturbation QG KE equation (\ref{qg_ke_eq}) in the $y-z$ plane for the case of $\beta=1.5$.}
%     \label{fig:qg_ke_beta1.5}
% \end{figure}
%%%%%%%%
%%%%%%%%

% %%%%%%
\clearpage
\bibliographystyle{jfm}
% Note the spaces between the initials
\bibliography{main}

\end{document}